\documentclass[manuscript, screen, nonacm]{acmart} 

\usepackage{amsmath}

\usepackage{breakcites}

\usepackage{dsfont}
  \usepackage{paralist}
  \usepackage{graphics} 
  \usepackage{epsfig} 
\usepackage{graphicx}  \usepackage{epstopdf}
 \usepackage{comment}
 \usepackage{enumitem}
 \usepackage{bm}
\usepackage{subcaption}
\usepackage{makebox}
\usepackage{booktabs}
\usepackage{todonotes}
\usepackage{appendix}
\usepackage{enumitem}
\usepackage{amsthm}

\usepackage{multirow}

\hypersetup{urlcolor=blue, citecolor=red}
\usepackage{tikz}
\usetikzlibrary{arrows.meta}
\usetikzlibrary{patterns}

\DeclareMathOperator\prox{prox}
\DeclareMathOperator*\argmin{argmin}

\newcommand{\codefont}{\textsf}

\newcommand{\cuqipy}{\codefont{CUQIpy}}
\newcommand{\vecd}[1]{\boldsymbol{#1}}
\newcommand{\lopd}[1]{\boldsymbol{#1}} 

\newcommand{\given}{\,\ifnum\currentgrouptype=16 \middle\fi|\,} 

\newcommand{\code}{\texttt}
\newcommand{\dd}{\mathrm{d}}

\usepackage{amsmath}
\DeclareMathOperator*{\argmax}{arg\,max}

\newcommand{\cuqitools}[1]{\textsf{CUQI} tools}

\usepackage[utf8]{inputenc}
\usepackage{pgfplots}
\DeclareUnicodeCharacter{2212}{−}
\usepgfplotslibrary{groupplots,dateplot}
\usetikzlibrary{patterns,shapes.arrows}
\pgfplotsset{compat=newest}

\usepackage{listings}

\definecolor{codegreen}{rgb}{0,0.6,0}
\definecolor{codegray}{rgb}{0.5,0.5,0.5}
\definecolor{codepurple}{rgb}{0.58,0,0.82}
\definecolor{backcolour}{rgb}{0.95,0.95,0.92}

\definecolor{cprior}{HTML}{CEF3D5}
\definecolor{clikelihood}{HTML}{FFF2C6}
\definecolor{cdata}{HTML}{BCD9FE}
\definecolor{cposterior}{HTML}{FFC6CA}
\definecolor{csampler}{HTML}{EAB5FE}
\definecolor{cstatistics}{HTML}{B9B9B9}

\lstdefinestyle{mystyle}{
    backgroundcolor=\color{backcolour},   
    commentstyle=\color{codegreen},
    keywordstyle=\color{magenta},
    numberstyle=\tiny\color{codegray},
    stringstyle=\color{codepurple},
    basicstyle=\ttfamily\footnotesize,
    breakatwhitespace=false,         
    breaklines=true,                 
    captionpos=b,                    
    keepspaces=true,                 
    numbers=left,                    
    numbersep=5pt,                  
    showspaces=false,                
    showstringspaces=false,
    showtabs=false,                  
    tabsize=2,
}

\usepackage{minted}
\definecolor{LightGray}{gray}{0.92}

\newenvironment{codebox}{\VerbatimEnvironment
\begin{minted}
[bgcolor=LightGray,mathescape]
{python}}
{\end{minted}}

\begin{document}

\title{A Computational Framework and Implementation of Implicit Priors in Bayesian Inverse Problems}

\author{Jasper M. Everink}
\email{jmev@dtu.dk}
\authornote{Both authors contributed equally to this research.}
\orcid{0000-0001-7263-0317}
\affiliation{%
  \institution{Technical University of Denmark}
  \city{Lyngby}
  \country{Denmark}
}

\author{Chao Zhang}
\email{chaz@dtu.dk}
\authornotemark[1]
\orcid{0000-0003-4402-3696}
\affiliation{%
  \institution{Technical University of Denmark}
  \city{Lyngby}
  \country{Denmark}
}

\author{Amal M. A. Alghamdi}
\email{amal.m.alghamdi@gmail.com}
\authornote{Currently at Impact Alpha, Saudi Arabia}
\orcid{0000-0003-0145-5296}
\affiliation{%
  \institution{Technical University of Denmark}
  \city{Lyngby}
  \country{Denmark}
}

\author{Rémi Laumont}
\email{remi.laumont@edf.fr}
\orcid{0009-0008-8364-7761}
\affiliation{%
    \institution{EDF R\&D}
    \city{Paris}
    \country{France}
}

\author{Nicolai A. B. Riis}
\orcid{0000-0002-6883-9078}
\affiliation{%
  \institution{Copenhagen Imaging ApS}
  \city{Lyngby}
  \country{Denmark}
}

\author{Jakob S. Jørgensen}
\email{jakj@dtu.dk}
\orcid{0000-0001-9114-754X}
\affiliation{%
  \institution{Technical University of Denmark}
  \city{Lyngby}
  \country{Denmark}
}

\renewcommand{\shortauthors}{Everink et al.}

\ccsdesc[500]{Applied computing~Mathematics and statistics}
\ccsdesc[300]{Mathematics of computing~Nonparametric statistics}

\keywords{Bayesian inverse problems, prior modelling, plug-and-play, regularization, sampling}

\begin{abstract}
Solving Bayesian inverse problems typically involves deriving a posterior distribution using Bayes' rule, followed by sampling from this posterior for analysis. Sampling methods, such as general-purpose Markov chain Monte Carlo (MCMC), are commonly used, but they require prior and likelihood densities to be explicitly provided. In cases where expressing the prior explicitly is challenging, implicit priors offer an alternative, encoding prior information indirectly. These priors have gained increased interest in recent years, with methods like Plug-and-Play (PnP) priors and Regularized Linear Randomize-then-Optimize (RLRTO) providing computationally efficient alternatives to standard MCMC algorithms. However, the abstract concept of implicit priors for Bayesian inverse problems is yet to be systematically explored and little effort has been made to unify different kinds of implicit priors. This paper presents a computational framework for implicit priors and their distinction from explicit priors. We also introduce an implementation of various implicit priors within the CUQIpy Python package for Computational Uncertainty Quantification in Inverse Problems. Using this implementation, we showcase several implicit prior techniques by applying them to a variety of different inverse problems from image processing to parameter estimation in partial differential equations.
\end{abstract}

\maketitle


\section{Introduction}
In the conventional approach to Bayesian inversion, a solution to an inverse problem is a posterior probability distribution on the parameter $\vecd{x}$ given observed measurements $\vecd{y}_{\text{obs}}$. Such a posterior probability distribution can be constructed using Bayes' rule, which in the setting of continuous probability distributions states that
\begin{equation}\label{eq:bayes}
    \underbrace{\pi(\vecd{x}|\vecd{y}_{\text{obs}})}_{\text{posterior}}\propto \underbrace{\pi(\vecd{y}_{\text{obs}}\,|\,\vecd{x})}_{\text{likelihood}}\underbrace{\pi(\vecd{x})}_{\text{prior}}.
\end{equation}
That is, the probability density of the posterior is proportional to the likelihood, which models the relation between the parameters $\vecd{x}$ and the data $\vecd{y}$, both of which are considered random variables, and a prior distribution, which models any information on the parameters $\vecd{x}$ in the absence of observed measurements $\vecd{y}_{\text{obs}}$.

If the posterior is not part of a well-understood family of probability distributions or the dimensionality is too large, it might not be feasible to analyse the posterior distribution directly. In this scenario, it is common to obtain a large number of samples from the posterior distribution and to use these samples for statistical analysis instead. One approach to sampling is by means of general-purpose Markov chain Monte Carlo (MCMC) algorithms, which requires the probability distribution, and therefore the likelihood and prior, to be specified by either supplying their densities $\pi(\vecd{y}_{\text{obs}}|\vecd{x})$ and $\pi(\vecd{x})$, or, for computational stability, their logarithms $\log\pi(\vecd{y}_{\text{obs}}|\vecd{x})$ and $\log\pi(\vecd{x})$. In the case of smooth potentials, there exists general-purpose algorithms, e.g., Langevin algorithms \cite{xifara2014langevin} and Hamiltonian Monte Carlo \cite{neal2011mcmc}, that make use of the potential gradients $-\nabla_{\vecd{x}} \log\pi(\vecd{y}_{\text{obs}}|\vecd{x})$ and $-\nabla_{\vecd{x}}\log\pi(\vecd{x})$. 

When we provide explicit expressions of the densities, by Bayes' rule, the posterior can also be written explicitly. We refer to this as an explicit approach with the priors being called \textbf{explicit priors}. However, sometimes we want to incorporate prior information that cannot be expressed as a density or accessing the density is intractable or undesirable. In this case we lack the explicit expression for the density and we refer to such priors as \textbf{implicit priors}. Fundamentally, the idea of implicit priors is a lack of certain desirable mathematical properties, but an increase in prior complexity and/or computational speed. Therefore, they lack a unified understanding which might be a reason the term is not consistently used in the literature for a single prior modelling strategy.

In parallel, explicit and implicit regularization is used within the deterministic setting. In particular within deep learning, where penalizing the objective function using weight decay/$l_2$ penalty and $l_1$ penalty are considered explicit regularization and techniques like early stopping and dropout are considered implicit regularization, see \cite{Goodfellow-et-al-2016}. Similarly, within inverse problems and statistics, penalties like Tikhonov regularization/Ridge regression/$l_2$ and LASSO/$l_1$ can be considered as explicit regularization, whilst iterative regularization can be considered as implicit regularization, see \cite{hansen2010discrete}.

Within the literature, the use of the terms explicit and implicit for probability distributions can slightly vary. Most common is to use implicit distribution to describe a probability distribution for which a density is intractable, but it can be analyzed by means of sampling, either exactly or approximately. Sometimes, the term is used when additional structure is imposed for convenience, e.g., being able to estimate expectations and its gradients \cite{huszar2017variational}, or the term is used for specific constructions, e.g., inverting data-generating statistical models \cite{diggle1984monte} or Fisher's fiducial argument \cite{mukhopadhyay2006some}.

Specialized to Bayesian statistics, the term ``implicit Bayes'' is used for the construction of posterior distributions without invoking or explicitly mentioning Bayes' rule \cite{cooper2013use}. However, within the machine learning community, it has also been associated with the concept of generative modelling \cite{mohamed2016learning}. In this setting, the prior is defined by some transformation $\vecd{x} = \lopd{G}(\vecd{z})$ of an auxiliary random variable $\vecd{z}$. Such transformation can be used to enforce constraints or sparsity \cite{xu2024bayesian}.  If the dimension of $\vecd{z}$ is smaller than the dimension of $\vecd{x}$ and $\lopd{G}$ is sufficiently regular, then $\vecd{x}$ lies on some low-dimensional space and the prior density of $\vecd{x}$ on this space might be infeasible to compute. In this scenario, it is preferred to study the posterior distribution on $\vecd{x}$ given $\vecd{y}$ through sampling from the posterior distribution on $\vecd{z}$ given $\vecd{y}$ followed by transforming the samples. Thus, although the posterior distribution on $\vecd{x}$ given $\vecd{y}$ does not have a tractable density, the auxiliary posterior distribution on $\vecd{z}$ given $\vecd{y}$ does have one that can be sampled from, and the prior distribution on $\vecd{x}$ is called implicit.

In this work, we also focus on two other classes of implicit priors that have appeared in recent years. One such prior can be constructed by directly incorporating prior information in the sampling algorithm instead of the density, similar to previously mentioned implicit regularization. This strategy contains the so called Plug-and-Play (PnP) priors. These PnP priors are generally characterized by replacing denoisers that appear in sampling algorithms from one derived from an explicit prior to an arbitrary denoiser, often one based on neural networks. Although originally proposed within the optimization context \cite{venkatakrishnan2013plug}, when applied to specific algorithms and denoisers, it can be shown that there does exist an underlying intractable prior, e.g., PnP-ULA \cite{laumont2022bayesian}, which we call an implicit prior.

Another recent class of implicit priors consists of incorporating prior information in another representation of the posterior distribution, instead of in the prior density. This approach contains the regularized linear Randomize-then-Optimize (RLRTO). This approach consists of enhancing a Gaussian distribution with the deterministic effects of regularization by adding explicit regularization to an optimization-based characterization of the posterior distribution. This allows for introducing constraints like nonnegativity \cite{bardsley2012mcmc} or sparsity \cite{everink2023sparse} in a computationally efficient manner, at the cost of making the posterior distribution and prior distribution implicit. For an in-depth comparison between this regularized approach and more conventional priors, see \cite{laumont2024sampling, zhang2025fast}.

Whilst all of the previously mentioned implicit priors have been applied in various contexts to solve inverse problems, there has not yet been an attempt at unification of the abstract concept of implicit priors. In this paper, we attempt this unification through the lens of a computational framework for implicit priors.  Furthermore, we are not aware of a single software package that can handle this variety of implicit priors. For example, \codefont{DeepInverse} \cite{tachella2025deepinverse}, a Python library for imaging with deep learning, only supports PnP-ULA out of the previously mentioned implicit priors. To this end, we provide an implementation of all previously mentioned implicit priors as part of the \cuqipy\,software package for Computational Uncertainty Quantification for Inverse Problems in Python \cite{riis2024cuqipy, alghamdi2024cuqipy}. 

\subsection{Contributions}
In this paper, we make the following contributions:
\begin{itemize}
    \item We present a computational framework for implicit priors that unifies generative models, Langevin algorithms, and RLRTO.
    \item We present an implementation of Langevin algorithms (ULA, MYULA and PnP-ULA) and RLRTO within the \cuqipy\, software package. Our implementation enables straightforward selection and application of implicit priors to complicated problems from applied mathematical and engineering fields. 

    \item We demonstrate the use of the implemented implicit priors to various inverse problems, including ones on which implicit priors have yet to be applied, e.g., PDE-based inverse problems, thereby showing the utility and flexibility of the abstract interpretation and numerical implementation.
\end{itemize}
  
\subsection{Overview}
This paper is structured as follows. Section \ref{sec:explicit} contains background on Bayesian inverse problems and commonly used explicit priors. Section \ref{sec:implicit} introduces the computational framework for implicit priors, discusses background on various implicit priors and relates these implicit priors to the presented computational framework. Section \ref{sec:cuqipy} contains a discussion of how to incorporate implicit priors within a computational Bayesian inverse problems framework, and showcases how to use the implicit priors discussed in Section \ref{sec:implicit} in \cuqipy. Section \ref{sec:examples} shows examples of implicit priors applied to solve various commonly considered inverse problems, namely: deblurring, PDE parameter estimation and image inpainting. Finally, Section \ref{sec:conclusion} contains some final concluding remarks.

\section{Bayesian inverse problems with explicit priors}\label{sec:explicit}

In this section, we summarize the basics of Bayesian inverse problems. We put particular emphasis on two commonly used families of prior distributions, namely uniform and Gaussian priors, as these provide simple examples that allow us to set up the notation and terminology necessary for later sections.

\subsection{Bayesian inverse problems}

The goal of (finite-dimensional) Bayesian inverse problems is to reconstruct a posterior probability distribution of a random variable $\vecd{x} \in \mathbb{R}^n$ representing the parameter of interest, given a random variable $\vecd{y} \in \mathbb{R}^m$ and an observation $\vecd{y}_{\text{obs}}\in \mathbb{R}^m$ representing the measurements. In one general form, the parameter $\vecd{x}$ and measurements $\vecd{y}$ are related through a forward operator $\lopd{A}: \mathbb{R}^n \rightarrow \mathbb{R}^m$ through the relation:
\begin{equation*}
    \vecd{y} = \lopd{F}(\lopd{A}(\vecd{x}), \vecd{e}),
\end{equation*}
where $\lopd{F}: \mathbb{R}^m \times \mathbb{R}^k \rightarrow \mathbb{R}^m$ corrupts $\lopd{A}(\vecd{x})$ with noise $\vecd{e} \in \mathbb{R}^k$. The most commonly studied form is additive noise of the form:
\begin{equation}\label{eq:linear_additive_model}
    \vecd{y} = \lopd{A}(\vecd{x}) + \vecd{e},
\end{equation}
where $\vecd{e} \in \mathbb{R}^m$ represents the noise and/or error, often modelled as a Gaussian distributed random variable.

To simplify the discussion, we will mainly consider a linear forward operator with additive Gaussian noise in examples and note that all examples, unless otherwise stated, can be generalized to non-linear forward operators and/or other noise models. In this case, we can describe \eqref{eq:linear_additive_model} by a probability distribution on the data, which for fixed observed measurements $\vecd{y}_{\text{obs}}$ we refer to as the likelihood function: 
\begin{equation}\label{eq:likelihood_function}
    L_{\vecd{y}_{\text{obs}}}(\vecd{x}) := \pi(\vecd{y}_{\text{obs}}\,|\,\vecd{x}) \propto \exp\left(-\frac{1}{2}\|\lopd{A}\vecd{x}-\vecd{y}_{\text{obs}}\|_{\lopd{\Sigma}_{\vecd{e}}^{-1}}^2\right),
\end{equation}
where we assume that the noise $\vecd{e}$ has mean zero and covariance $\lopd{\Sigma}_{\vecd{e}}\in \mathbb{R}^{m\times m}_{>0}$ and $\|\mathbf{z}\|_{\lopd{M}}^2 = \mathbf{z}^T \lopd{M} \mathbf{z}$ denotes the squared weighted norm.

To obtain a probability distribution on $\vecd{x}$ given observed measurements $\vecd{y}_{\text{obs}}$, we start out with a prior distribution $\pi(\vecd{x})$, followed by an application of Bayes' rule to obtain a posterior distribution:
\begin{equation}\label{eq:bayes_formula}
    \pi(\vecd{x}\,|\,\vecd{y}_{\text{obs}}) \propto L_{\vecd{y}_{\text{obs}}}(\vecd{x})\pi(\vecd{x}).
\end{equation}
For example, a commonly chosen family of prior distributions takes the form:
\begin{equation}\label{eq:prior}
    \pi(\vecd{x}) \propto \exp\left(-R(\vecd{x})\right),
\end{equation}
where $R: \mathbb{R}^n \rightarrow \mathbb{R}$ is the called the prior potential. After updating prior \eqref{eq:prior} with likelihood \eqref{eq:likelihood_function} by applying Bayes' rule, the posterior takes the form:
\begin{equation}\label{eq:posterior}
    \pi(\vecd{x}\,|\,\vecd{y}_{\text{obs}}) \propto \exp\left(-\frac{1}{2}\|\lopd{A}\vecd{x}-\vecd{y}_{\text{obs}}\|_{\lopd{\Sigma}_{\vecd{e}}^{-1}}^2-R(\vecd{x})\right).
\end{equation}

A prior of the form \eqref{eq:prior} is commonly chosen, because its associated maximum a posteriori (MAP) estimate is equivalent to a regularized least squares problem, i.e.,
\begin{equation}\label{eq:MAP}
    \argmax_{\vecd{x}}\left\{\pi(\vecd{x}\given \vecd{y}_{\text{obs}})\right\} = \argmin_{\vecd{x}}\left\{\frac{1}{2}\|\lopd{A}\vecd{x}-\vecd{y}_{\text{obs}}\|_{\lopd{\Sigma}_{\vecd{e}}^{-1}}^2+R(\vecd{x})\right\}.
\end{equation}
In particular, when the prior is a Gaussian distribution, this corresponds to ridge regression/Tikhonov regularization, whilst if the prior is a Laplace distribution, this corresponds to the LASSO, thus resulting in explicit regularization.

Once a posterior distribution has been constructed, it is often analysed by computing samples from the distribution. For certain posterior distributions there exist specialized sampling algorithms, whilst otherwise more general families of sampling methods are used, for example, the Langevin methods which will be discussed in Subsection \ref{subsec:langevin}. Finally, empirical summary statistics are computed using these samples, often including a point estimator, e.g., MAP, mean or componentwise median, together with an estimate of the uncertainty, e.g., standard deviation or credible interval widths.

\subsection{Explicit priors}\label{ssec:explicit_priors}
For a direct application of Bayes' rule \eqref{eq:bayes_formula}, we need to provide an expression of the prior density $\pi(\vecd{x})$. We call priors for which we specify the prior density \textbf{explicit}. To illustrate these explicit expressions and discuss some related technical considerations, we briefly discuss two simple families of prior distributions, namely uniform and Gaussian priors. Similar considerations can be made about other, more general families of prior distributions, including non-informative priors (e.g., Jeffreys prior) and sparsity-promoting and edge-preserving priors (e.g., Laplace \cite{bardsley2012laplace} and Cauchy priors \cite{markkanen2019cauchy}).

\subsubsection{Uniform priors}
In a uniform prior, any permissible parameter is assigned equal weight. More formally, let $\Omega \subsetneq \mathbb{R}^n$ be some bounded, measurable domain of permissible parameters. The uniform prior on $\Omega$ takes the form:
\begin{equation}\label{eq:bounded_uniform_prior}
    \pi(\vecd{x}) = \frac{1}{|\Omega|}\mathbb{I}\{\vecd{x}\in \Omega\} := \begin{cases}
        \frac{1}{|\Omega|},\quad &\text{if } \vecd{x} \in \Omega, \text{ and}\\
        0,\quad &\text{otherwise,}
    \end{cases}
\end{equation}
where $|\Omega| = \int_{\Omega}1\text{d}\vecd{x}$ is the volume of $\Omega$.

Applying Bayes' rule results in the posterior distribution:
\begin{equation}\label{eq:uniform_posterior}
    \pi(\vecd{x}\given \vecd{y}_{\text{obs}}) \propto  L_{\vecd{y}_{\text{obs}}}(\vecd{x}) \pi(x) \propto  \exp\left(-\frac{1}{2}\|\lopd{A}\vecd{x}-\vecd{y}_{\text{obs}}\|_{\lopd{\Sigma}_{\vecd{e}}^{-1}}^2 \right)\mathbb{I}\{\vecd{x}\in \Omega\}.
\end{equation}
We assume boundedness of $\Omega$ such that $|\Omega| < \infty$, in which case posterior \eqref{eq:uniform_posterior} is guaranteed to be a probability distribution, and we can therefore sample from it. However, it is not uncommon to choose $\Omega$ such that $|\Omega| = \infty$, e.g., the unconstrained setting $\Omega = \mathbb{R}^n$, in which case \eqref{eq:bounded_uniform_prior} is not properly defined. In this case, we can still define a uniform prior, even though it will not correspond to an actual probability distribution. Such a prior will take the form:
\begin{equation}\label{eq:unbounded_uniform_prior}
    \pi(\vecd{x}) = \mathbb{I}\{\vecd{x}\in \Omega\}.
\end{equation}

Even though \eqref{eq:unbounded_uniform_prior} might not be normalizable, it is common to abuse notation and not make a distinction and write $\pi(\vecd{x}) \propto \mathbb{I}\{\vecd{x}\in \Omega\}$ for any measurable $\Omega$. Similarly, after applying Bayes' rule, the posterior takes again the form of \eqref{eq:uniform_posterior}. However, because the prior does not correspond to a probability distribution, it is not guaranteed that the posterior does so as well. For convenience, we call a posterior \textbf{proper} if it corresponds to an actual probability distribution, from which one can sample, and \textbf{improper} otherwise. Following \cite{gelman1995bayesian}, we call a prior \textbf{proper} if it corresponds to an actual probability distribution and does not depend on the observed measurements $\vecd{y}_{\text{obs}}$. This second condition excludes priors constructed using empirical Bayes from being proper. For a concrete example, if we consider a uniform prior on $\Omega = \mathbb{R}^n$, then the prior is improper. It can be easily shown that the associated posterior distribution \eqref{eq:uniform_posterior} is proper if and only if $\text{nullspace}(\lopd{A}) = \{\vecd{0}\}$, which does not hold for underdetermined systems of equations. A simple computational example of such a setting can be found in Subsection~\ref{subsec:cp_explicit}.

\subsubsection{Gaussian priors}
Another popular prior for Bayesian inverse problems is the Gaussian prior, due to its easy interpretation and computational efficiency. A Gaussian prior is defined by a mean vector $\vecd{\mu} \in \mathbb{R}^n$ and a symmetric positive definite covariance matrix $\lopd{\Sigma}_p \in \mathbb{R}^{n\times n}_{++}$ which together specify a Gaussian prior of the form:
\begin{equation}\label{eq:Gaussian_prior}
    \pi(\vecd{x}) \propto \exp\left(-\frac{1}{2}\|\vecd{x}-\vecd{\mu}\|_{\lopd{\Sigma}_{p}^{-1}}^2\right).
\end{equation}

After applying Bayes' rule we end up with the posterior:
\begin{equation}\label{eq:Gaussian_posterior}
    \pi(\vecd{x}\given \vecd{y}_{\text{obs}}) \propto L_{\vecd{y}_{\text{obs}}}(\vecd{x}) \pi(x) \propto  \exp\left(-\frac{1}{2}\|\lopd{A}\vecd{x}-\vecd{y}_{\text{obs}}\|_{\lopd{\Sigma}_{\vecd{e}}^{-1}}^2 -\frac{1}{2}\|\vecd{x}-\vecd{\mu}\|_{\lopd{\Sigma}_{p}^{-1}}^2\right).
\end{equation}

Instead of specifying the covariance matrix $\lopd{\Sigma}_{p}$, it is also common to specify the precision matrix $\lopd{\Lambda} = \lopd{\Sigma}_{p}^{-1}$ or a square root of the precision matrix, i.e., $\lopd{\Gamma} \in \mathbb{R}^{r\times n}$ such that $\lopd{\Lambda} = \lopd{\Gamma}^T\lopd{\Gamma} \in \mathbb{R}^{n\times n}_{++}$.

The symmetric positive definiteness assumption on the covariance matrix, or equivalently the precision matrix, guarantees that posterior \eqref{eq:Gaussian_posterior} is proper. However, it is not uncommon that specifying $\lopd{\Gamma}$ such that $\lopd{\Lambda}$ is only symmetric positive semidefinite, e.g., if $r < n$. In this case, is can be easily shown that prior \eqref{eq:Gaussian_prior} is improper whilst posterior \eqref{eq:Gaussian_posterior} is only proper if $\text{nullspace}(\lopd{A})\cap\text{nullspace}(\lopd{\Gamma}) = \{\vecd{0}\}$. A simple computational example of such a setting can be found in Subsection \ref{subsec:cp_explicit}.

\section{Bayesian inverse problems with implicit priors}\label{sec:implicit}
In this previous section we have seen two families of prior distributions, namely Gaussian and Uniform distributions, that we can specify explicitly by providing the prior density $\pi(\vecd{x})$. Whilst providing the density explicitly is beneficial for being able to use many different sampling algorithms, e.g., Metropolis-Hastings algorithms, many different sampling algorithms do not directly make use of $\pi(\vecd{x})$, but another quantity derived from $\pi(\vecd{x})$. 

The simplest example is implementing Metropolis-Hastings algorithms using logarithmic densities. In this case, one implements $\log \pi(\vecd{x})$ instead of $\pi(\vecd{x})$ to improve the numerical stability of the algorithms. In this relatively simple case, computing the prior density $\pi(\vecd{x})$ from its logarithm $\log \pi(\vecd{x})$ can be done efficiently by applying the exponential function. But many other sampling algorithms make use of quantities that are derived from $\pi(\vecd{x})$, whilst $\pi(\vecd{x})$ cannot be efficiently computed from the derived quantity or there might be no theoretical guarantee that such a $\pi(\vecd{x})$ even exists. This leads us to assuming the following definition of \textbf{implicit priors}:
\begin{center}
\textit{
prior information provided to a sampling algorithm in the case that the corresponding explicit prior density is either expensive to compute or non-existent.}
\end{center}

This definition of implicit priors is inherently subjective as we do not formalize what we mean by computationally expensive. We consider the operation of applying the exponential function to $\log \pi(\vecd{x})$ as inexpensive, and discuss the rest on a case-by-case basis.

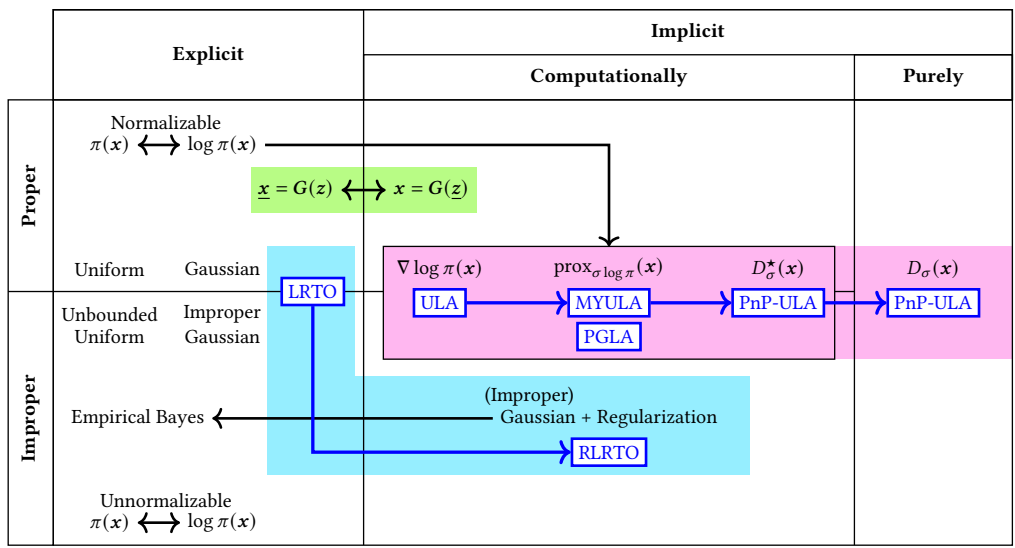
\begin{figure}[tb]
    \centering
    \tikzstyle{tline}=[black, line width = 0.5]
\tikzstyle{tarrow}=[black, line width = 1]
\tikzstyle{tdashed}=[dashed, black, line width = 1.25]
\tikzstyle{ttext}=[black, line width = 1]
\tikzstyle{talgo}=[blue, draw=blue, line width = 1]
\tikzstyle{talgoarrow}=[blue, draw=blue, line width = 1.25]

\begin{tikzpicture}[scale = 0.75, every node/.style={scale=0.85}]
    \definecolor{clatent}{rgb}{0.722, 0.988, 0.522};
    \definecolor{crto}{rgb}{0.588, 0.933, 1};
    \definecolor{clangevin}{rgb}{1, 0.718, 0.941};

    \def\width{17.0};
    \def\expb{5.5}
    \def\pureb{2.8}
    
    \def\height{-9.5};
    \def\himprop{-4.5}
    
    \def\hunit{-0.4};
    \def\mainh{4*\hunit};
    \def\wunit{1};

    \def\langhalfwidth{4};
    \def\langhalfheight{2*\hunit}


    \fill[fill = crto] (\expb - 1.3*\wunit-0.4*\wunit,\height-\himprop-2*\hunit) rectangle (0.5*\expb + 0.5*\width-0.5*\pureb + 2.5*\wunit, \height-0.5*\himprop + 2.5*\hunit);
    \fill[fill = white] (0.5*\expb + 0.5*\width-0.5*\pureb - \langhalfwidth - 0.5*\wunit, \height-\himprop - \langhalfheight) rectangle (\width, \height-\himprop + 1.5*\langhalfheight - 0.3);
    
    \fill[fill = clangevin] (0.5*\expb + 0.5*\width-0.5*\pureb - \langhalfwidth, \height-\himprop - \langhalfheight) rectangle (\width, \height-\himprop + 1.5*\langhalfheight);
    
    \fill[fill = clatent] (\expb - 2*\wunit, \mainh + 3*\hunit) rectangle (\expb + 2*\wunit, \mainh + 5*\hunit);

    \draw[tline] (0, 0) rectangle (\width, \height);
    \draw[tline] (0, 0) rectangle (\width, 4*\hunit);
    \draw[tline] (0, 0) rectangle (\expb, 4*\hunit);
    \draw[tline] (\expb, 2*\hunit) rectangle (\width, 4*\hunit);
    \draw[tline] (\width-\pureb, 2*\hunit) rectangle (\width, 4*\hunit);
    \draw[tline] (\expb, 4*\hunit) -- (\expb,\height);
    \draw[tline] (\width-\pureb, 4*\hunit) -- (\width-\pureb,\height);
    \draw[tline] (2*\hunit,\height-\himprop) -- (\width-\pureb,\height-\himprop);
    
    \draw[tline] (2*\hunit, 4*\hunit) rectangle (0, \height);

    \node[ttext] at (0.5*\expb,2*\hunit) {\textbf{Explicit}};
    \node[ttext] at (0.5*\expb + 0.5*\width,1*\hunit) {\textbf{Implicit}};
    \node[ttext] at (0.5*\expb + 0.5*\width-0.5*\pureb,3*\hunit) {\textbf{Computationally}};
    \node[ttext] at (\width-0.5*\pureb,3*\hunit) {\textbf{Purely}};

    \node[ttext, rotate=90] at (1*\hunit,0.5*\height+2*\hunit-0.5*\himprop) {\textbf{Proper}};
    \node[ttext, rotate=90] at (1*\hunit,\height-0.5*\himprop) {\textbf{Improper}};

    \node[ttext] () at (2*\wunit,\mainh + \hunit) {Normalizable};
    \node[ttext] (density_proper) at (\wunit,\mainh + 2*\hunit) {$\pi(\vecd{x})$};
    \node[ttext] (logdensity_proper) at (3*\wunit,\mainh + 2*\hunit) {$\log\pi(\vecd{x})$};
    \draw[tarrow, <->] (density_proper) -- (logdensity_proper);
    
    \node[ttext] () at (2*\wunit,\height - 2*\hunit) {Unnormalizable};
    \node[ttext] (density_improper) at (\wunit,\height - 1*\hunit) {$\pi(\vecd{x})$};
    \node[ttext] (logdensity_improper) at (3*\wunit,\height - 1*\hunit) {$\log\pi(\vecd{x})$};
    \draw[tarrow, <->] (density_improper) -- (logdensity_improper);

    \node[ttext] (latent_x) at (\expb - 1.2*\wunit, \mainh + 4*\hunit) {$\underline{\vecd{x}} = \lopd{G}(\vecd{z})$};
    \node[ttext] (latent_z) at (\expb + 1.2*\wunit, \mainh + 4*\hunit) {$\vecd{x} = \lopd{G}(\underline{\vecd{z}})$};
    \draw[tarrow, <->] (latent_x) -- (latent_z);

    \node[ttext] () at (\wunit,\height-\himprop-1*\hunit) {Uniform};
    \node[ttext] () at (\wunit,\height-\himprop+1*\hunit) {Unbounded};
    \node[ttext] () at (\wunit,\height-\himprop+2*\hunit) {Uniform};
    
    \node[ttext] () at (3*\wunit,\height-\himprop-1*\hunit) {Gaussian};
    \node[ttext] () at (3*\wunit,\height-\himprop+1*\hunit) {Improper};
    \node[ttext] () at (3*\wunit,\height-\himprop+2*\hunit) {Gaussian};

    \node[ttext] (EB) at (\expb - 4*\wunit, \height-0.5*\himprop) {Empirical Bayes};

    \filldraw[tline, fill = clangevin] (0.5*\expb + 0.5*\width-0.5*\pureb - \langhalfwidth, \height-\himprop - \langhalfheight) rectangle (0.5*\expb + 0.5*\width-0.5*\pureb + \langhalfwidth, \height-\himprop + 1.5*\langhalfheight);

    \node[ttext] (gradient) at (0.5*\expb + 0.5*\width-0.5*\pureb - 3*\wunit, \height-\himprop - \hunit) {$\nabla\log\pi(\vecd{x})$};
    \node[talgo, fill = white] (ULA) at (0.5*\expb + 0.5*\width-0.5*\pureb - 3*\wunit, \height-\himprop + 0.5*\hunit) {ULA};
    
    \node[ttext] (prox) at (0.5*\expb + 0.5*\width-0.5*\pureb, \height-\himprop - \hunit) {$\text{prox}_{\sigma\log\pi}(\vecd{x})$};
    \node[talgo, fill = white] (MYULA) at (0.5*\expb + 0.5*\width-0.5*\pureb, \height-\himprop + 0.5*\hunit) {MYULA};
    \node[talgo, fill = white] () at (0.5*\expb + 0.5*\width-0.5*\pureb, \height-\himprop + 2*\hunit) {PGLA};
    
    \node[ttext] (MMSE) at (0.5*\expb + 0.5*\width-0.5*\pureb + 3*\wunit, \height-\himprop - \hunit) {$D_\sigma^\star(\vecd{x})$};
    \node[talgo, fill = white] (PnP_ULA_1) at (0.5*\expb + 0.5*\width-0.5*\pureb + 3*\wunit, \height-\himprop + 0.5*\hunit) {PnP-ULA};
    
    \node[ttext] (non_MMSE) at (\width - 0.5*\pureb , \height-\himprop - \hunit) {$D_\sigma(\vecd{x})$};
    \node[talgo, fill = white] (PnP_ULA_2) at (\width - 0.5*\pureb, \height-\himprop + 0.5*\hunit) {PnP-ULA};

    \draw[talgoarrow, ->] (ULA) -- (MYULA);
    \draw[talgoarrow, ->] (MYULA) -- (PnP_ULA_1);
    \draw[talgoarrow, ->] (PnP_ULA_1) -- (PnP_ULA_2);
    
    \draw[tarrow, ->] (logdensity_proper) -- (0.5*\expb + 0.5*\width-0.5*\pureb, \mainh + 2*\hunit) -- (0.5*\expb + 0.5*\width-0.5*\pureb, \mainh + 6.5*\hunit);

    \node[ttext] () at (0.5*\expb + 0.5*\width-0.5*\pureb - 1.4\wunit, \height-0.5*\himprop - \hunit) {(Improper)};
    \node[ttext] (RLRTO_data) at (0.5*\expb + 0.5*\width-0.5*\pureb , \height-0.5*\himprop) {Gaussian + Regularization};
    \draw[tarrow, ->] (RLRTO_data) -- (EB);

    \node[talgo, fill = white] (LRTO) at (\expb - 0.9*\wunit,\height-\himprop) {LRTO};
    \node[talgo, fill = white] (RLRTO) at (0.5*\expb + 0.5*\width-0.5*\pureb, \height-0.5*\himprop + 1.5*\hunit) {RLRTO};
    
    \draw[talgoarrow, ->] (LRTO)  -- (\expb - 0.9*\wunit, \height-0.5*\himprop + 1.5*\hunit) -- (RLRTO);
    
\end{tikzpicture}
    
    \caption{Classification of different prior information based on proper/improper and explicit/implicit. Arrows describe how the head of the arrow can in many cases be derived from the tail of the arrow. When multiple inferable variables occur, the focus is on the underlined variable. Blue boxes represent different sampling algorithms. The light green area represents latent variable and generative model-based priors discussed in Subsection~\ref{subsec:latvar}. The pink area represents Langevin methods discussed in Subsection~\ref{subsec:langevin}. The light blue area represents Randomize-then-Optimize based methods discussed in Subsection~\ref{subsec:rto}.}
    \label{fig:prior_zoo}
\end{figure}

Our continued discussion within this section is summarized within Figure~\ref{fig:prior_zoo}, which contains a structured overview of the implicit priors we will consider. We will summarize various existing methods and discuss their interpretation as implicit priors. Specifically, in Subsection~\ref{subsec:latvar} we will discuss latent variable and generative models, which provide a first look into implicit priors and what we might consider as computationally expensive. Then in Subsection~\ref{subsec:langevin}, we discuss Langevin methods and PnP, which use various quantities derived from the prior density with various levels of difficulty of computing the corresponding explicit prior. The Langevin methods also include cases where no explicit prior needs to exist. Finally, in Subsection \ref{subsec:rto}, we discuss the RLRTO method and its associated prior.

\subsection{Latent variable and generative models}\label{subsec:latvar}
\begin{figure}[tb]
\definecolor{matplot_blue}{HTML}{1f77b4}
\definecolor{matplot_orange}{HTML}{ff7f0e}
\definecolor{matplot_green}{HTML}{2ca02c}
    \centering
    \begin{subfigure}{0.49\textwidth}
        \begin{tikzpicture}
        \node[] at (0,0) {$\lopd{G}$};
        \draw[->] (-0.9,-0.5) -- (0.9,-0.5);
        
        \node[] at (-2,-2) {$\vecd{z} \in \mathbb{R}$};
        \draw[] (-2,-1.5) -- (-2,1);
        
        \node[] at (2,-2) {$\vecd{x} \in \mathbb{R}^2$};
        \draw[] plot [smooth, tension=1] coordinates {(2,-1.5) (1.5,-0.5) (2.5,0.5) (2,1)};
    \end{tikzpicture}
    \end{subfigure}%
    \begin{subfigure}{0.49\textwidth}
    \begin{tikzpicture}
        \node[] at (0,0) {$\lopd{G}$};
        \draw[->] (-0.9,-0.5) -- (0.9,-0.5);

        \node[] at (-2,-2) {$\vecd{z} \in \mathbb{R}^2$};
        \filldraw[matplot_blue] (-2,0.5) circle (2pt);
        \filldraw[matplot_orange] (-1.5,-0.5) circle (2pt);
        
        \node[] at (2,-2) {$\vecd{x} \in \mathbb{R}^n$};
        
        \draw[matplot_blue] (1,-1) -- (2,-1) -- (2, 0.5) -- (3, 0.5);
        \draw[matplot_orange] (1,-0.5) -- (2,-0.5) -- (2, 0) -- (3, 0);

    \end{tikzpicture}
    \end{subfigure}%
    \caption{Two examples of change of variables resulting in priors implicit in $\vecd{x}$, but explicit in $\vecd{z}$. On the left: a mapping from a simple space into a more complicated one, e.g., by means of a generative model. On the right: a mapping form a low-dimensional space onto piecewise-constant signals in a higher-dimensional space.}
    \label{fig:change_of_variables}
\end{figure}
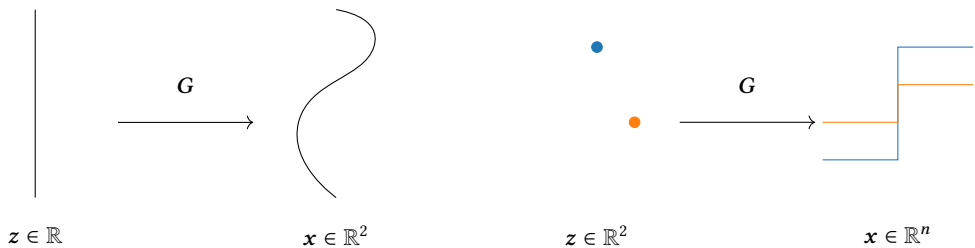

One alternative approach to describing priors is by means of transforming some auxiliary random variable into the parameter of interest. This approach is referred to as a latent variable or generative model. Formally, such a prior is constructed by transforming an auxiliary random variable $\vecd{z} \in \mathbb{R}^k$ into the parameter $\vecd{x} \in \mathbb{R}^n$ using some transformation $\lopd{G}: \mathbb{R}^k \rightarrow \mathbb{R}^n$. Often, the auxiliary variable is assumed to be distributed according to a standard multivariate normal distribution, such that the prior takes the form:
\begin{align}
    \vecd{x} &= \lopd{G}(\vecd{z}), \text{ with} \label{eq:transform}\\
    \pi(\vecd{z}) &\propto \exp\left(-\frac12\|\vecd{z}\|_2^2\right).\label{eq:auxiliary}
\end{align}
To sample from \eqref{eq:transform}, we first compute a standard multivariate normal sample $\vecd{z}$ and then apply $\lopd{G}$ to obtain a sample from $\vecd{x}$. The function $\lopd{G}$ could be used for various reasons: a change of variables to make sampling more efficient, to enforce constraints by mapping to an unconstrained distribution onto the constraint set, or as a generative model like auto-decoders \cite{kingma2013auto, laloy2017inversion}, normalizing flows \cite{whang2021composing} and diffusion processes \cite{chung2022diffusion}. See Figure \ref{fig:change_of_variables} for some simple illustrative examples.

However, because the prior distribution on $\vecd{x}$ is not specified by means of a density $\pi(\vecd{x})$, we cannot directly apply Bayes' formula. Dependent on the choice of $\lopd{G}$, various approaches can be used to still incorporate this prior in the Bayesian framework. In this work, we focus on relatively simple transformations of variables and low-dimensional latent spaces to simplify the exposition.

\subsubsection{Change of variables}
Consider the case where $\lopd{G}: \mathbb{R}^n \rightarrow \mathbb{R}^n$ is bijective with a differentiable inverse. In this case we can apply a change of variables rule for probability densities to explicitly write down a density for $\vecd{x}$. More specifically we get:
\begin{equation}\label{eq:change_variable_explicit}
    \pi(\vecd{x}) \propto \left|\frac{\text{d}\lopd{G}^{-1}}{\text{d}\vecd{x}}(\vecd{x})\right|\exp\left(-\frac12\|\lopd{G}^{-1}(\vecd{x})\|_2^2\right).
\end{equation}
For example, this can be used to enforce positivity by using the componentwise transformation $\lopd{G}(\vecd{z})_i = \exp(\vecd{z}_i)$ for $i = 1,\dots,n$. This is a common technique for enforcing positivity in parameter estimation of partial differential equations \cite{dashti2015bayesian}.

From the perspective of explicit versus implicit priors we have two cases. Either \eqref{eq:change_variable_explicit} can be efficiently computed, in which case we can use the explicit prior density $\pi(\vecd{x})$, or \eqref{eq:change_variable_explicit} cannot be efficiently computed, in which case using $\pi(\vecd{x})$ for sampling is not efficient and implicit methods might have to be considered.

\subsubsection{No change of variables}
A change of variables formula like \eqref{eq:change_variable_explicit} is not always available or practical. Consider for example the transformation $\lopd{G}(\vecd{z})_i = \max(\vecd{z}_i, 0)$ for $i = 1,\dots,n$. This transformation guarantees $\vecd{x}$ to be nonnegative, but is not invertible. Furthermore, samples from $\vecd{z}$ with at least one negative component gets mapped to the boundary of the set of nonnegative vectors. Thus, $\vecd{x}$ lies on this boundary with positive probability, which cannot be described using a continuous probability density $\pi(\vecd{x})$ making sampling more difficult. Similarly, if $\lopd{G}: \mathbb{R}^k \rightarrow \mathbb{R}^n$ with $k < n$, as is the case with for example auto-decoders, then $\vecd{x}$ lies on the lower-dimensional image of $\lopd{G}$, and thus $\vecd{x}$ does not have a density on $\mathbb{R}^n$. Whilst a density on this lower-dimensional image might exist, computing it for a complicated transformation $\lopd{G}$ is not practical. In both of these examples, whilst we can sometimes compute the explicit prior density $\pi(\vecd{x})$, either on the whole of $\mathbb{R}^n$ or a subspace, this is not the most computationally efficient way for sampling.

Instead of focusing on directly inferring $\vecd{x}$, we can move a step back and try to infer $\vecd{z}$, followed by afterwards transforming any samples of $\vecd{z}$ into $\vecd{x}$. We are then able to write $L_{\vecd{y}_{\text{obs}}}(\vecd{x}) = L_{v\vecd{y}_{\text{obs}}}(\lopd{G}(\vecd{z}))$, allowing us to apply Bayes' rule on $\vecd{z}$ resulting in the posterior:
\begin{equation}\label{eq:auxiliary_posterior}
    \pi(\vecd{z}\given \vecd{y}_{\text{obs}}) \propto  L_{\vecd{y}_{\text{obs}}}(\lopd{G}(\vecd{z}))\pi(\vecd{z}) \propto \exp\left(-\frac{1}{2}\|\lopd{A}\lopd{G}(\vecd{z})-\vecd{y}_{\text{obs}}\|_{\lopd{\Sigma}_{\vecd{e}}^{-1}}^2 - \frac12\|\vecd{z}\|_2^2\right).
\end{equation}
In this case we could consider the prior to be explicit in the auxiliary variable $\vecd{z}$ even though it only implicitly described the prior density on parameter of interest $\vecd{x}$. Note that although we now have a posterior probability density to sample from, writing the posterior as \eqref{eq:auxiliary_posterior} can increase the complexity of sampling, with a possibly linear forward $A$ being replaced by the now often non-linear function $\lopd{A} \circ \lopd{G}$.

\subsection{Langevin methods}\label{subsec:langevin}
In order to sample from a high-dimensional target posterior, a popular
and important class of methods is based on the Langevin stochastic differential equation (SDE), which for posterior sampling takes the form:
\begin{equation}\label{eq:langevin_sde}
    \dd \vecd{x}_t = \nabla \log \pi(\vecd{x}_t\given \vecd{y}_{\text{obs}}) dt + \sqrt{2}\dd\vecd{b}_t,
\end{equation}
with Brownian motion $(\vecd{b}_t)_{t\geq 0}$. If $\pi(\cdot\given \vecd{y}_{\text{obs}})$ is proper and smooth, and $\nabla \log \pi(\cdot\given \vecd{y}_{\text{obs}})$
is $L$-Lipschitz continuous, then for any initial condition $\vecd{x}_0$, the Langevin SDE
\eqref{eq:langevin_sde} admits a unique solution $(\vecd{x}_t)_{t\geq 0}$ with the target posterior as unique stationary density \cite{roberts1996exponential}. Thus, to sample from the posterior, we can sample from a solution of the Langevin SDE.

\subsubsection{Unadjusted Langevin Algorithms}
However, solving \eqref{eq:langevin_sde} exactly is rarely possible in practice and we need to discretize
\eqref{eq:langevin_sde} in order to approximately sample from our target posterior. Various discretizations for SDEs can be used, with one particularly popular scheme being the Euler-Maruyama discretization, which scales well with the dimension of the problem and is relatively straightforward to apply \cite{durmus2017nonasymptotic}. This discretization leads to the Unadjusted Langevin Algorithm (ULA), which takes the form:
\begin{equation}\label{eq:ula}
    \vecd{x}_{k+1} = \vecd{x}_k + \delta \nabla \log L_{\vecd{y}_{\text{obs}}}(\vecd{x}_k) + \delta \nabla \log \pi (\vecd{x}_k) + \sqrt{2\delta}\vecd{z}_{k+1},
\end{equation}
with $(\vecd{z}_k)_{k \in \mathbb{N}}$ a sequence of iid Gaussian random variables with zero mean and identity covariance and a discretization step-size $0<\delta <1/L$.

It can be helpful to interpret ULA as stochastic gradient-descent \cite{welling2011bayesian}. Similar to the gradient-descent scheme, $\delta$ controls the trade-off between the discretization error and the convergence speed. For example, a large step size allows us to rapidly explore the parameter-space and might converge faster, but is more prone to instabilities and comes with more asymptotic bias. For a small enough $\delta$, the ULA Markov chain admits a unique stationary distribution close to the posterior \cite{roberts1996exponential}. In addition non-asymptotic error bounds between this stationary density and the target density allows us to quantify the distance between the two distributions \cite{durmus2017nonasymptotic}.

When sampling from a posterior distribution using ULA, \eqref{eq:ula} shows that
we need both the logarithm of the prior density and of the likelihood to be differentiable. Note in particular that we do not need an explicit expression of the prior $\pi(\vecd{x})$, but only the gradient of its logarithm, i.e., $\nabla \log \pi (\vecd{x})$, thus we could consider the case of providing this gradient as an implicit prior. Consequently, whilst we could use ULA to sample from posteriors with Poisson likelihood or with priors that are almost everywhere differentiable, such as the total-variation prior or the Laplace prior, it may result in samples that do not properly represent the posterior distribution.

Whilst we only need to provide the gradient $\nabla \log \pi (\vecd{x})$, we could in theory solve a differential equation of the form $f = \nabla \phi$ to compute $\log \pi (\vecd{x})$ up to an additive constant and then apply the exponential function to compute $\pi (\vecd{x})$. However, the computational cost of being able to solve the differential equation at an arbitrary point $\vecd{x}$ might not be worth the effort to compensate the small approximation error caused by the discretization of the Langevin SDE. One way to remove this approximation error is by using ULA as the proposal distribution in a Metropolis-Hastings algorithm, the so called Metropolis Adjusted Langevin Algorithm (MALA), but this again requires either $\log \pi (\vecd{x})$ or $\pi (\vecd{x})$ and is therefore only practical when available in a computationally efficient form.

Whilst using $\nabla \log \pi (\vecd{x})$ as an implicit prior, one could sample from the corresponding prior by using ULA \eqref{eq:ula} without the likelihood term and assuming that $\pi (\vecd{x})$ is proper. Thus whether this implicit prior is proper or not completely depends on whether the underlying explicit prior is proper or not. 

An alternative discretization scheme, based on forward-backward splitting of the likelihood and prior term, is the proximal gradient Langevin algorithm (PGLA), see \cite{salim2020primal, ehrhardt2024proximal}. PGLA takes the form:
\begin{equation}\label{eq:PGLA}
    \vecd{x}_{k+1} = \prox_{\delta \log \pi (\cdot)}\left(\vecd{x}_k + \delta \nabla \log L_{\vecd{y}_{\text{obs}}}(\vecd{x}_k) + \sqrt{2\delta}\vecd{z}_{k+1}\right),
\end{equation}
with $(\vecd{z}_k)_{k \in \mathbb{N}}$ and $\delta$ the same as for ULA \eqref{eq:ula}, and  $\prox$ denotes the proximal operator defined by:
\begin{equation}\label{eq:proximal}
    \prox_{\alpha R}(\vecd{x}) := \argmin_{\vecd{z}}\left\{\frac12\|\vecd{x}-\vecd{z}\|_2^2 + \alpha R(\vecd{z})\right\}.
\end{equation}

Unlike ULA \eqref{eq:ula}, PGLA \eqref{eq:PGLA} makes use of the proximal operator of the log prior instead of the gradient of the prior logarithm. For many priors, proximal operator can either be computed through a closed form expression or an efficient algorithm to approximate it. Therefore, it can be beneficial to use $\prox_{\delta \log \pi (\cdot)}$ as an implicit prior; whilst it might be theoretically possible to compute the prior density $\log \pi(\vecd{x})$ from the associated proximal operator, this is not a commonly occuring computational problem.

Various other unadjusted Langevin algorithms based on splitting that do not use the (log) prior directly exist. Examples include subgradient Langevin methods like Prox-sub and Grad-sub \cite{habring2024subgradient}, and ULPDA \cite{narnhofer2024posterior}.

\subsubsection{Moreau-Yoshida Unadjusted Langevin Algorithm (MYULA)}
We now assume that the prior takes the form $\pi(\vecd{x})
\propto \exp(-R(\vecd{x}))$ with $R$ proper, lower semi-continuous and convex, but generally not differentiable everywhere. In this case, ULA \eqref{eq:ula} might not be suitable. In order to overcome the non-differentiability issue, it was proposed in \cite{durmus2018efficient} to apply ULA to a surrogate, smooth posterior
distribution $\pi_{\alpha}(\vecd{x}\given\vecd{y}_{\text{obs}}) \propto  L_{\vecd{y}_{\text{obs}}}(\vecd{x}) \pi_\alpha(\vecd{x})$ with $\pi_\alpha(\vecd{x}) \propto
\exp (-R_\alpha (\vecd{x}))$. The potential $R_\alpha$ stands for the Moreau-Yoshida envelope
of $R$ which is the minimum value at which the proximal operator \eqref{eq:proximal} is attained and is defined for all $\vecd{x} \in \mathbb{R}^d$ by
\begin{equation}\label{eq:moreau_envelope}
    R_\alpha (\vecd{x}) = \operatorname{inf}_{\vecd{z}} \left\{\frac{1}{2}\|\vecd{x}-\vecd{z} \|_2^2 + \alpha R(\vecd{z})\right\}.
\end{equation}

For sampling from the surrogate smooth posterior by ULA, we need  the gradient of the Moreau-Yoshida envelope $R_\alpha$, which satisfies:
\begin{equation}\label{eq:grad_prox}
     -\nabla \log \pi_\alpha(\vecd{x}) = \nabla R_\alpha (\vecd{x}) = \frac{1}{\alpha}\left(\text{prox}_{\alpha R}(\vecd{x}) - \vecd{x}\right).
\end{equation}

Combining \eqref{eq:grad_prox} with ULA \eqref{eq:ula} results in the Moreau-Yoshida Unadjusted Langevin Algorithm (MYULA):
\begin{equation}\label{eq:my_ula}
    \vecd{x}_{k+1} = \vecd{x}_k + \delta \nabla \log L_{\vecd{y}_{\text{obs}}}(\vecd{x}_k) + \frac{\delta}{\alpha}\left(\text{prox}_{\alpha R}(\vecd{x}_k) - \vecd{x}_k\right) +
    \sqrt{2\delta} \vecd{z}_{k+1}.
\end{equation}
Similarly to PGLA \eqref{eq:PGLA}, we do not provide the density $\pi(\vecd{x})
\propto \exp(-R(\vecd{x}))$ explicitly, but only describe it implicitly through the proximal operator $\prox_{\alpha R}(\vecd{z})$ and can thus be considered as an implicit prior. Simple examples of using MYULA are shown in Subsection \ref{subsec:langevin_in_cuqipy}.

\subsubsection{Plug-and-Play priors}
There are various approaches to incorporating neural networks in solving inverse problems, see \cite{habring2024neural} for an overview. One such approach, which is very suitable for sampling, consists of the so called Plug-and-Play (PnP) priors, which work as follows.

Note that the proximal operator $\text{prox}_{\alpha R}$ defined in \eqref{eq:proximal} solves a denoising problem with iid Gaussian noise with zero mean and variance $\alpha$ using a regularized linear least squares problem. Thus, in every iteration of MYULA \eqref{eq:my_ula} or PGLA \eqref{eq:PGLA} the iterate is denoised by the proximal operator induced by the log prior. A natural generalization is to replace the proximal operator by an arbitrary denoiser, henceforth referred to as a \textbf{restorator}. Replacing a proximal operator $\text{prox}_{\alpha R}$ induced by a prior with an arbitrary restorator $D_{\alpha}$ is referred to as Plug-and-Play, due to how simple it is to switch between restorators and obtain experimental results.

Such a PnP approach can be applied to any algorithm which uses the proximal operator of a prior term, not necessarily sampling. For example, in optimization, combining PnP with a proximal gradient (ISTA) method results in PnP-ISTA \cite{meinhardt2017learning} and combining PnP with ADMM results in PnP-ADMM \cite{chan2016plug}.

Our main example for sampling is its application to MYULA, resulting in PnP-ULA \cite{laumont2022bayesian}. Replacing the proximal in \eqref{eq:my_ula} by a restorator $D_{\alpha}$ results in:
\begin{equation}\label{eq:pnp_ula}
    \vecd{x}_{k+1} = \vecd{x}_k + \delta \nabla \log L_{b}(\vecd{x}_k) + \frac{\delta}{\alpha}\left(D_{\alpha}(\vecd{x}_k) - \vecd{x}_k\right) +
    \sqrt{2\delta} \vecd{z}_{k+1} \ .
\end{equation}
Though replacing the proximal with an arbitrary restorator can feel practically sane, theoretical study is required to show that it actually targets a posterior distribution and that the algorithm actually converges. Without any assumptions on $D_\alpha$, we cannot guarantee that \eqref{eq:pnp_ula} samples from an actual probability distribution, in this case we cannot speak of a proper posterior distribution, let alone a prior distribution. In this case, whilst we provide implicit prior information, the prior information exists purely implicitly due the lack of existence of an actual prior. However, as discussed in \cite{laumont2022bayesian}, if the restorator is a minimum mean square error (MMSE) estimator $D^\star_{\alpha}$, then by Tweedie's identity there does indeed exist a prior $\pi(\vecd{x})$ from which this restorator can be derived. However, additional conditions are needed to guarantee convergence of the PnP-ULA algorithm and whether we can correctly sample from the prior and/or posterior. But in all cases, obtaining an explicit expression for the prior, if it exists, is intractable, hence all PnP priors are implicit. 

\subsection{Randomize-then-Optimize methods}\label{subsec:rto}
Another approach to sampling is by means of Randomize-then-Optimize (RTO) based methods. Randomize-then-Optimize \cite{bardsley2014randomize}, as originally proposed for sampling from posteriors with non-linear forward operators, is a Metropolis-Hastings algorithm where the proposal distribution is defined by means of a randomized optimization problem that is comparable to MAP estimation. This approach originated as a generalization of the linear forward operator setting, referred to by us as linear Randomize-then-Optimize (LRTO), discussed next. LRTO has furthermore been extended to incorporate regularization, which we refer to as regularized linear Randomize-then-Optimize (RLRTO) and the implicit prior technique we consider. Simple examples of LRTO and RLRTO are shown in Subsection \ref{subsec:cp_RLRTO}.

\subsubsection{Linear Randomize-then-Optimize (LRTO)}
Recall that in the case of a linear forward operator with additive Gaussian noise and a Gaussian prior, the posterior takes the form:
\begin{equation}\label{eq:posterior_gaussian}
    \pi(\vecd{x}\given \vecd{y}_{\text{obs}})\propto \pi(\vecd{y}_{\text{obs}}\given \vecd{x})\pi(\vecd{x})\propto \exp\left(-\frac12\|\lopd{A}\vecd{x}-\vecd{y}_{\text{obs}}\|^2_{\lopd{\Sigma}_e^{-1}}-\frac12\|\vecd{x}-\vecd{\mu}\|^2_{\lopd{\Sigma}_p^{-1}}\right).\\
\end{equation}

It can be shown that this posterior distribution is itself again a Gaussian distribution, a property known as conjugacy. This favourable property can be exploited to sample efficiently from the posterior. One such method is LRTO, which repeatedly solves randomized linear least squares problems. Specifically, it characterizes the Gaussian posterior as the random solution to the following randomized optimization problem:
\begin{align}
    &\argmin_{\vecd{x} \in \mathbb{R}^n} \big\{\underbrace{\frac12\|\lopd{A}\vecd{x}-\hat{\vecd{y}}\|^2_{\lopd{\Sigma}_e^{-1}}}_{\text{likelihood}} + \underbrace{\frac12\|\vecd{x}-\vecd{\hat{\mu}}\|^2_{\lopd{\Sigma}_p^{-1}}}_{\text{prior}}\big\},\ \label{eq:linear_rto} \\
    &\text{with}\ \hat{\vecd{y}} \sim \mathcal{N}(\vecd{y}_{\text{obs}}, \lopd{\Sigma}_e) \text{ and } \hat{\vecd{\mu}} \sim \mathcal{N}(\vecd{\mu}, \lopd{\Sigma}_p). \nonumber
\end{align}
To obtain an independent sample from posterior \eqref{eq:posterior_gaussian}, we sample independently from the perturbed data $\hat{\vecd{y}}$ and prior mean $\hat{\vecd{\mu}}$, which can often be done very efficiently, fill them into the linear least squares equation \eqref{eq:linear_rto}, and solve the optimization problem to a suitable accuracy. Due to techniques from numerical linear algebra, this can result in a very efficient algorithm to sample from Gaussian posterior distributions.

\subsubsection{Regularization Linear Randomize-then-Optimize (RLRTO)}
The characterization of the posterior with the randomized optimization problem like \eqref{eq:linear_rto} is limited to the explicit Gaussian priors, both proper and improper. To still make use of efficient solvers for optimization problems, it was proposed in \cite{everink2023sparse} to incorporate information directly into the optimization problem instead of the prior density by adding regularization to the randomized linear least squares problem \eqref{eq:linear_rto}. The result is the following randomized optimization problem: 
\begin{align}
    &\argmin_{\vecd{x} \in \mathbb{R}^n} \big\{\underbrace{\frac12\|\lopd{A}\vecd{x}-\hat{\vecd{y}}\|^2_{\lopd{\Sigma}_e^{-1}}}_{\text{likelihood}} + \underbrace{\frac12\|\vecd{x}-\hat{\vecd{\mu}}\|^2_{\lopd{\Sigma}_p^{-1}} + R(\vecd{x})}_{\text{prior}}\big\},\label{eq:regularized_linear_rto}\\
    &\text{with}\ \hat{\vecd{y}} \sim \mathcal{N}(\vecd{y}_{\text{obs}}, \lopd{\Sigma}_e) \text{ and } \hat{\vecd{\mu}} \sim \mathcal{N}(\vecd{\mu}, \lopd{\Sigma}_p). \nonumber
\end{align}

In \cite{everink2023sparse}, the regularization term $R(\vecd{x})$ is assumed to be convex, with particular emphasis on the sparsity-promoting regularization, like $l_1$ penalties, and constraints, like nonnegativity. For these regularization terms, it was shown that samples from \eqref{eq:regularized_linear_rto} are sparse with positive probability.

Just like for the randomized linear least squares problem \eqref{eq:linear_rto}, a randomized optimization problem does not define a sampler. To turn it into a sampler, we need to repeatedly solve \eqref{eq:regularized_linear_rto} using some optimization algorithm. The choice of optimization algorithm greatly influences the speed of the sampling. As proposed in \cite{Everink2023, everink2023sparse, zhang2025fast}, we could use algorithms like L-BFGS-B, FISTA or ADMM, depending on the chosen regularization term. 

It has been empirically shown in \cite{zhang2025fast} that using RLRTO can be faster than other implicit priors that promote true sparsity, i.e., samples with zero coefficients. Alternatively, modelling true sparsity explicitly using a non-continuous prior distribution is an option, but the corresponding sampling algorithm like RJMCMC and STMALA are notorious for not scaling well to high-dimensional problems, see \cite{schreck2015shrinkage}.

Note that the prior information is incorporated implicitly in optimization problem \eqref{eq:regularized_linear_rto} by means of the Gaussian prior mean $\vecd{\mu}$, Gaussian prior covariance $\lopd{\Sigma}_{p}$, and the regularization function $R(\vecd{x})$, and not in an explicit density for $\pi(\vecd{x})$. To guarantee that the solution of \eqref{eq:regularized_linear_rto} is unique for every instance of $\hat{\vecd{y}}$ and $\hat{\vecd{\mu}}$, we can choose $\lopd{\Sigma}^{-1}_{p}$ to be positive definite. Showing uniqueness of \eqref{eq:regularized_linear_rto} in the case where $\lopd{A}$ has a non-trivial null space and $\lopd{\Sigma}_{p}^{-1}$ is not positive definite is a computationally difficult problem in itself, see \cite{everink2024geometry}.

Understanding what the explicit prior distribution looks like requires some mathematical effort. As mentioned in \cite{Everink2023, everink2023sparse, astfalck2018posterior}, it can be shown that there does indeed exist a prior distribution that corresponds the posterior distribution defined by \eqref{eq:regularized_linear_rto}. However, as shown in \cite{Everink2023, everink2023sparse}, computing this prior distribution would require computing high-dimensional Gaussian integrals over polyhedral sets, which is computationally expensive and hence we consider RLRTO as an implicit prior. Furthermore, this prior distribution generally depends on the observed measurements $\vecd{y}_{\text{obs}}$ and is therefore improper. Whilst we could therefore consider RLRTO as an instance of empirical Bayes, accessing the empirical Bayesian prior is not feasible.

To give some examples of the complicated nature of prior distribution resulting from RLRTO, let $R(\vecd{x}) = \chi_{K}(\vecd{x})$, where $K$ denotes an affine subspace and $\chi_{K}(\vecd{x})$ is $0$ if $\vecd{x} \in K$ and infinite otherwise. In this special scenario the prior corresponds to taking a Gaussian prior $\vecd{x} \sim \mathcal{N}(\vecd{\mu}, \lopd{\Sigma}_p)$ and restricting it to the subspace $K$. In cases where the regularization function $R(\vecd{x})$ promotes sparsity, e.g., $R(\vecd{x}) = \|\vecd{x}\|_1$, the posterior assigns positive probability to $\vecd{0}$. Similarly, the posterior assigns for every subset of coefficients of $\vecd{x}$ a positive probability to the event that $\vecd{x}$ has zeroes in that subset of coefficients. Thus, this posterior distribution assigns positive probability to any of exponentially many subsets of $\mathbb{R}^n$ of varying dimension.

\section{Implicit priors in \cuqipy}\label{sec:cuqipy}
In this Section, we discuss how to use implicit priors within the \cuqipy\, software package for Computational Uncertainty Quantification for Inverse Problems in Python. 
We first provide a general discussion and design considerations on incorporating implicit priors within a computational Bayesian workflow. This is then followed a large number of examples of how to use implicit priors discussed in Section \ref{sec:implicit} within \cuqipy. These examples consider solving two very simple linear and non-linear inverse problems with two parameters and a single measurement. Examples of solving more complicated inverse problems using these methods are shown in Section \ref{sec:examples}, though no code will be shown there. All code examples can be found at \href{https://doi.org/10.5281/zenodo.17115255}{https://doi.org/10.5281/zenodo.17115255}.

\subsection{Implicit priors in a computational Bayesian workflow}

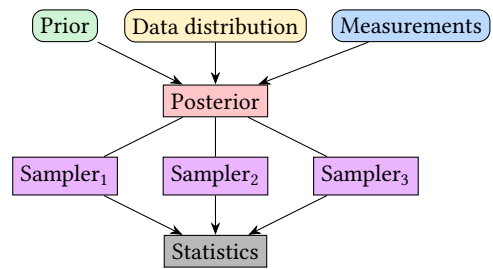
\begin{figure}[tb]
    \centering
    \begin{tikzpicture}[
>={Stealth[black]},
every node/.style={draw=black, shape=rectangle, minimum size=1em},
every edge/.style={draw=black, very thick}
]
    \node[draw, rounded corners=.15cm, fill=cprior] (prior) at (-2,-1) {Prior};
    \node[draw, rounded corners=.15cm, fill=clikelihood] (likelihood) at (0, -1) {Data distribution};
    \node[draw, rounded corners=.15cm, fill=cdata] (data) at (2.6, -1) {Measurements};
    \node[draw, fill=cposterior] (posterior) at (0,-2.0) {Posterior};

    \node[draw, fill=csampler] (sampler1) at (-2,-3) {Sampler\textsubscript{1}};
    \node[draw, fill=csampler] (sampler2) at (0,-3) {Sampler\textsubscript{2}};
    \node[draw, fill=csampler] (sampler3) at (2,-3) {Sampler\textsubscript{3}};
    
    \node[draw, fill=cstatistics] (statistics) at (0,-4) {Statistics};

    \draw[->] (prior) -- (posterior);
    \draw[->] (likelihood) -- (posterior);
    \draw[->] (data) -- (posterior);
    \draw[->] (posterior) -- (sampler1) -- (statistics);
    \draw[->] (posterior) -- (sampler2) -- (statistics);
    \draw[->] (posterior) -- (sampler3) -- (statistics);
\end{tikzpicture}
    \caption{A standard Bayesian workflow as used in \cuqipy.}
    \label{fig:Bayesian_workflow_standard}
\end{figure}

A general computational Bayesian workflow can look as follows: 
\begin{enumerate}
    \item Define a prior on $\vecd{x}$ that describes all the information we have about the unknown parameter.
    \item Define the data distribution, which describes the probability of the measurements $\vecd{y}$ given parameters $\vecd{x}$. In combination with the observed data $\vecd{y}_{obs}$, one obtains the likelihood from the data distribution.
    \item Combine the prior, data distribution and the observed measurements $\vecd{y}_{\text{obs}}$ into a posterior distribution.
    \item Sample from the posterior using one of many samplers to obtain samples.
    \item Summarize the posterior using sample statistics.
\end{enumerate}
This workflow is illustrated in Figure~\ref{fig:Bayesian_workflow_standard}, in which different samplers should result in approximations of the same posterior statistics.

Implementing the various implicit priors considered in Section \ref{sec:implicit} as Python classes into this workflow comes with varying levels of difficulty. Latent variables can be implemented rather easily by incorporating the transformation into the data distribution such that the posterior is defined over the auxiliary variable. The resulting posterior samples can then be transformed back into the parameter of interest before computing any sample statistics.

The Langevin methods based on the gradient of the log prior, the proximal of the log prior, or an MMSE denoiser can also be implemented with relative ease, as they are all derived from explicit prior densities and the choice of sampler merely introduces an approximation error.

However, general restorators with PnP-ULA method and the RLRTO approach require more considerations. Neither implicit priors are derived from an explicit prior, but both are defined by incorporating the implicit information directly into another representation of a posterior distribution, either a Langevin algorithm or a randomized optimization problem. This implicit prior information is therefore introduced in the later stages of the Bayesian workflow, in particular in the sampler stage. In this stage we consider the randomized optimization problem characterization of a Gaussian posterior already a sampler, making no distinction between the choice of optimizer.

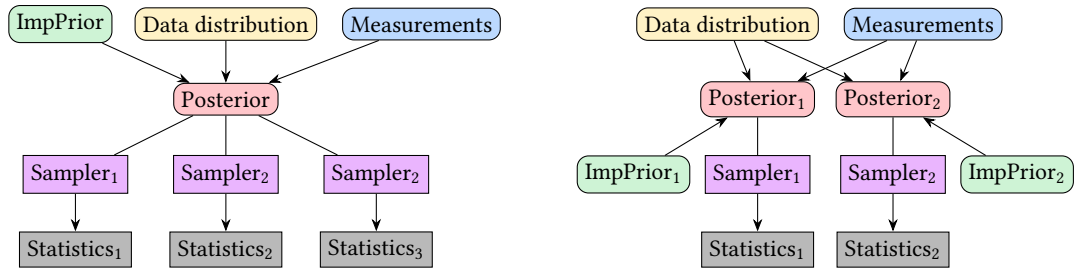
\begin{figure}[tb]
    \centering
    \begin{subfigure}{0.5\textwidth}
        \begin{tikzpicture}[
>={Stealth[black]},
every node/.style={draw=black, shape=rectangle, minimum size=1em},
every edge/.style={draw=black, very thick}
]
    
    \node[draw, rounded corners=.15cm, fill=cprior] (imp_prior) at (-2.2,0) {ImpPrior};
    \node[draw, rounded corners=.15cm, fill=clikelihood] (likelihood) at (0, 0) {Data distribution};
    \node[draw, rounded corners=.15cm, fill=cdata] (data) at (2.6, 0) {Measurements};
    
    \node[draw, rounded corners=.15cm, fill=cposterior] (posterior) at (0,-1.0) {Posterior};

    \node[draw, fill=csampler] (sampler1) at (-2,-2) {Sampler\textsubscript{1}};
    \node[draw, fill=csampler] (sampler2) at (0,-2) {Sampler\textsubscript{2}};
    \node[draw, fill=csampler] (sampler3) at (2,-2) {Sampler\textsubscript{2}};
    
    \node[draw, fill=cstatistics] (statistics1) at (-2,-3) {Statistics\textsubscript{1}};
    \node[draw, fill=cstatistics] (statistics2) at (0,-3) {Statistics\textsubscript{2}};
    \node[draw, fill=cstatistics] (statistics3) at (2,-3) {Statistics\textsubscript{3}};

    \draw[->] (imp_prior) -- (posterior);
    \draw[->] (likelihood) -- (posterior);
    \draw[->] (data) -- (posterior);
    \draw[->] (posterior) -- (sampler1) -- (statistics1);
    \draw[->] (posterior) -- (sampler2) -- (statistics2);
    \draw[->] (posterior) -- (sampler3) -- (statistics3);
\end{tikzpicture}
    \end{subfigure}%
    \begin{subfigure}{0.5\textwidth}
        \begin{tikzpicture}[
>={Stealth[black]},
every node/.style={draw=black, shape=rectangle, minimum size=1em},
every edge/.style={draw=black, very thick}
]

    \node[draw, rounded corners=.15cm, fill=cprior] (imp_prior1) at (-1.55,-2) {ImpPrior\textsubscript{1}};
    \node[draw, rounded corners=.15cm, fill=cprior] (imp_prior2) at (3.55,-2) {ImpPrior\textsubscript{2}};
    
    \node[draw, rounded corners=.15cm, fill=clikelihood] (likelihood) at (-0.3, 0) {Data distribution};
    \node[draw, rounded corners=.15cm, fill=cdata] (data) at (2.3, 0) {Measurements};
    
    \node[draw, rounded corners=.15cm, fill=cposterior] (posterior1) at (0.1,-1.0) {Posterior\textsubscript{1}};
    \node[draw, rounded corners=.15cm, fill=cposterior] (posterior2) at (1.9,-1.0) {Posterior\textsubscript{2}};

    \node[draw, fill=csampler] (sampler1) at (0.1,-2) {Sampler\textsubscript{1}};
    \node[draw, fill=csampler] (sampler2) at (1.9,-2) {Sampler\textsubscript{2}};
    
    \node[draw, fill=cstatistics] (statistics1) at (0.1,-3) {Statistics\textsubscript{1}};
    \node[draw, fill=cstatistics] (statistics2) at (1.9,-3) {Statistics\textsubscript{2}};

    \draw[->] (imp_prior1) -- (posterior1);
    \draw[->] (likelihood) -- (posterior1);
    \draw[->] (posterior1) -- (sampler1) -- (statistics1);

    \draw[->] (imp_prior2) -- (posterior2);
    \draw[->] (likelihood) -- (posterior2);
    \draw[->] (posterior2) -- (sampler2) -- (statistics2);

    \draw[->] (data) -- (posterior1);
    \draw[->] (data) -- (posterior2);
\end{tikzpicture}
    \end{subfigure}%
    \caption{Implicit priors within a standard Bayesian workflow, either considering the choice of sampler as separate from the implicit prior (left) or as a part of the implicit prior (right).}
    \label{fig:Bayesian_workflow_implicit}
\end{figure}

This leaves us with the choice whether to make the sampler a part of the implicit prior in the Bayesian workflow:
\begin{itemize}
    \item Consider the samplers as not being part of the implicit prior class, this is illustrated in the left side of Figure~\ref{fig:Bayesian_workflow_implicit}. In this case, the implicit prior class only stores the information needed for a sampler to use as an implicit prior, e.g., the restorator or regularization, greatly reducing the number of implicit prior classes. However, once a posterior is made using an implicit prior, it need not specify a probability distribution yet in the absence of a sampler. This posterior could then be sampled from using different samplers, which could result in drastically different sample statistics. One could consider a single implicit prior class that stores a restorator, e.g., an MMSE denoiser, a proximal operator of a prior logarithm or a proximal operator of a regularization function. Once this implicit prior is combined into a posterior, it could be sampled from by various samplers, including MYULA/PnP-ULA and RLRTO, with a solver that uses the proximal operator of the regularization, such as FISTA. Naturally, these approaches are very different and will result in very different sample statistics.
    \item Alternatively, consider making the sampler a part of the implicit prior class, this is illustrated in the right side of Figure~\ref{fig:Bayesian_workflow_implicit}. The main benefit of this approach is that once the posterior is defined, it corresponds to an actual probability distribution and any sampler that accepts this posterior, even though only one may be implemented, results in approximations of the same posterior statistics. This would imply that every new PnP algorithm that can be studied in the future requires its own implicit prior class, even though they could all be using the same restorator. Yet this would make sense for the RLRTO approach, in which the choice of regularization is naturally considered to be part of the prior information and not of the sampler.
\end{itemize}

Note that neither of these approaches are perfect for all implicit priors considered and careful consideration has to be made to make it fit naturally within the Bayesian workflow. Hence, we decided for \cuqipy{} to use a combination of both. More concretely, we have made the following decisions:
\begin{itemize}
    \item We have decided to emply different approaches for Langevin-based and RLRTO methods.
    \item For Langevin-based methods, we use a single base implicit prior class called restoration prior which can store an arbitrary restorator. Once made part of a posterior, any sampler that only needs the posterior gradient can be used, with the restorator accessed using \eqref{eq:grad_prox}, with the proximal operator replaced by the restorator. We therefore disregard the possible differences caused by not using a MMSE denoiser with different PnP algorithms, as this is a commonly used assumption with PnP algorithms.
    \item For RLRTO, we use a single base implicit prior class, which we refer to as a regularized Gaussian prior. This implicit prior contains the information of a Gaussian and the regularization term; once made part of a posterior, it can only be sampled from using an RLRTO sampler, which can choose the optimization algorithm based on the regularization term provided.
\end{itemize}

In the remainder of this section, we show how the various explicit and implicit priors discussed in Section \ref{sec:explicit} and \ref{sec:implicit}, which are summarized in Figure~\ref{fig:prior_zoo}, can be used within \cuqipy, based on the considerations discussed in this subsection.

\subsection{Explicit priors in \cuqipy}\label{subsec:cp_explicit}
Consider ``probably the simplest Bayesian inverse problem in the world'' from \cite{cuqibook}, which is a linear inverse problem that consists of two unknown parameters and a single measurement:
\begin{equation}\label{eq:simplest_linear}
    y = x_1 + x_2 + e, \text{ with } e \sim \mathcal{N}(0, 0.1),
\end{equation}
with ground truth $x_1 = x_2 = 1.5$ and observed measurement $y_{\text{obs}} = 3.0338$. Within \cuqipy\,, this simple linear problem consists of a \texttt{LinearModel} representing the forward operator and the observed measurement \texttt{y\textunderscore obs}:

\begin{codebox}
A = LinearModel(np.array([[1.0, 1.0]]))
y_obs = np.array([3.0338])
\end{codebox}
In the above and every future codebox, \texttt{np} is the \texttt{numpy} Python package.

Now to show the first example of how to solve Bayesian inverse problem \eqref{eq:simplest_linear}, we consider the following two equations:
\begin{align}
    \vecd{x} &\sim \mathcal{N}(\vecd{0}, 10\lopd{I})\label{eq:example_prior_model}, \text{ and}\\
    \vecd{y} &\sim \mathcal{N}(\lopd{A}\vecd{x}, 0.1\lopd{I})\label{eq:example_data_model},
\end{align}
where \eqref{eq:example_prior_model} represents a Gaussian prior and \eqref{eq:example_data_model} represents the data distribution \eqref{eq:simplest_linear} and thereby the likelihood. Equations \eqref{eq:example_prior_model} and \eqref{eq:example_data_model} can be translated directly into \cuqipy:
\begin{codebox}
x = Gaussian(np.zeros(2), 10)
y = Gaussian(A@x, 0.1)
\end{codebox}
Note the almost direct translation of the mathematical equations into Python code.

Next, we create the posterior distribution associated with \eqref{eq:example_prior_model} and \eqref{eq:example_data_model} by combining the random variables into a joint distribution on $(\vecd{x}, \vecd{y})$, followed by conditioning on the observed data \texttt{y\textunderscore obs}:
\begin{codebox}
posterior = JointDistribution(x, y)(y=y_obs)
\end{codebox}

Finally, we can sample from the posterior using one of many sampling algorithms, e.g., using a random walk Metropolis-Hastings, which requires the logarithm of the prior density to be implemented:
\begin{codebox}
samples = MH(posterior, scale=0.5).sample(10000).get_samples()
\end{codebox}
Every hundredth sample from this posterior distribution are shown in the top-left of Figure~\ref{fig:simplest_linear}.

If we know a priori that $x_1$ and $x_2$ are at least zero and never greater than two, we could use a uniform prior distribution $\vecd{x} \sim \text{Unif}(0, 2)^2$ instead, resulting in the code:
\begin{codebox}
x = Uniform(np.zeros(2), 2*np.ones(2))
y = Gaussian(A(x), 0.1)
posterior = JointDistribution(x, y)(y=y_obs)
\end{codebox}
Samples from the posterior distribution associated with this uniform prior are shown in the top-right of Figure~\ref{fig:simplest_linear}.

\subsection{Latent variables}

Assume that we know a priori that both $x_1$ and $x_2$ are strictly positive. One way of realizing this is by designing a prior on $(x_1, x_2)$ with support on $(0, \infty)^2$. Examples would be the product of two gamma or lognormal distributions. Consider the latter, which is defined to be the exponential of a single-variate Gaussian distribution. We could therefore construct a lognormal distribution by using a transformation $\vecd{x} = \lopd{G}(\vecd{z}) = \exp(\vecd{z})$ applied to a Gaussian distributed latent variable $\vecd{z}$. Similar to the previous example, we therefore consider the componentwise exponentiation:
\begin{align*}
    \lopd{G}(\vecd{z}) &:= \exp(\vecd{z}), \text{ and}\\
    \lopd{A}_{\text{exp}}(\vecd{z}) &:= \lopd{A}\lopd{G}(\vecd{z}).
\end{align*}

\begin{figure}[tb]
    \centering
    \begin{subfigure}{0.33\textwidth}
        \centering
        \includegraphics[width=\textwidth]{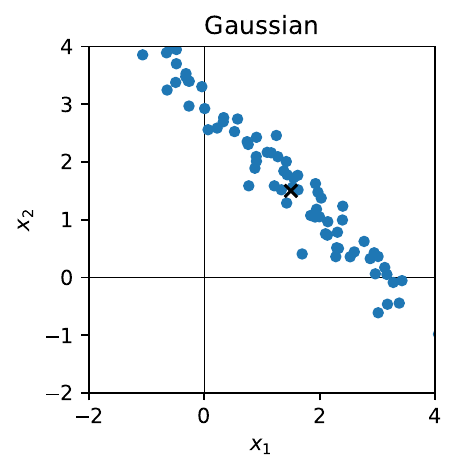}
    \end{subfigure}%
    \begin{subfigure}{0.33\textwidth}
        \centering
        \includegraphics[width=\textwidth]{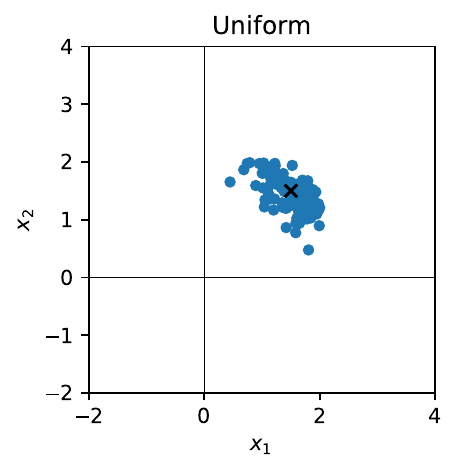}
    \end{subfigure}\\
    \begin{subfigure}{0.33\textwidth}
        \centering
        \includegraphics[width=\textwidth]{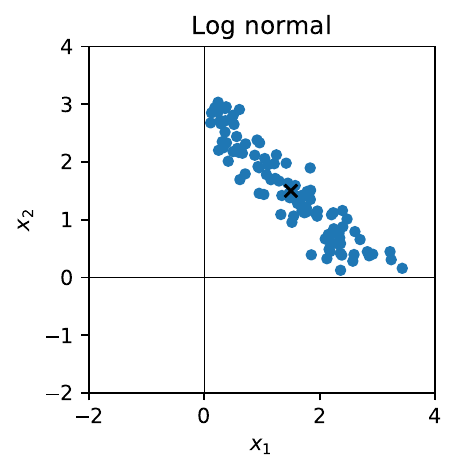}
    \end{subfigure}%
    \begin{subfigure}{0.33\textwidth}
        \centering
        \includegraphics[width=\textwidth]{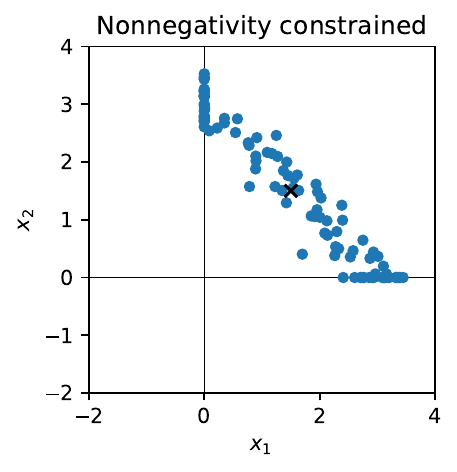}
    \end{subfigure}
    
    \caption{Samples of posteriors for the simple linear inverse problem \eqref{eq:simplest_linear} using various priors: explicit Gaussian and uniform, latent variable through exponential transformation, and RLRTO with nonnegativity.}
    \label{fig:simplest_linear}
\end{figure}

To do this in \cuqipy{}, we could either incorporate $\lopd{G}$ in the forward model ourselves, or make use of \texttt{Geometry} objects. In \cuqipy{}, geometries describe the space on which (random) variables are defined. It always contains the dimension of the space, but can also describe the interpretation. For example, a \texttt{Continuous1D} geometry describes a discretized, one-dimensional signal, whilst a \texttt{Discrete} geometry describes parameters without any connectivity. This additional descriptor makes it easier to visualize various posterior statistics. Besides this functionality, it can also be used to describe a transformation of parameters. Within \cuqipy{}, such a transformation can be achieved using \texttt{MappedGeometry}. The corresponding \cuqipy{} code looks as follows: 
\begin{codebox}
geom = MappedGeometry(Discrete(2), lambda z : np.exp(z))
z = Gaussian(np.zeros(2), 1, geometry=geom)
A_exp = LinearModel(np.array([[1.0, 1.0]]), domain_geometry=geom)
y = Gaussian(A_exp(z), 0.1)
\end{codebox}
Samples from the corresponding posterior are shown in the bottom-left of Figure~\ref{fig:simplest_linear}.

\subsection{RLRTO}\label{subsec:cp_RLRTO}
Now consider again Gaussian prior \eqref{eq:example_prior_model}. With the linear forward model and additive Gaussian noise assumption, the posterior distribution is again a Gaussian and thus LRTO \eqref{eq:linear_rto} can be applied. Within \cuqipy, this sampling algorithm is simply called \texttt{LinearRTO} and is applied as follows:
\begin{codebox}
samples = LinearRTO(posterior).sample(100)
\end{codebox}

In the above code, the sampler precisely formulates the randomized optimization problem \eqref{eq:linear_rto}, and solves it approximately using a conjugate gradient least squares (CGLS) algorithm. 

In order to incorporate constraints and regularization using RLRTO, we have introduced the \texttt{RegularizedGaussian} class to represent the implicit prior. Distributions from the \texttt{RegularizedGaussian} class work just like the Gaussian distribution, with the addition of additional parameters to specify the constraint and/or regularization. Thus, if we know a priori that both $x_1$ and $x_2$ are nonnegative, we can simply specify nonnegativity constraints to the Gaussian at the top of page 16 as follows:
\begin{codebox}
x = RegularizedGaussian(np.zeros(2), 10,
                        constraint="nonnegativity")
\end{codebox}

Using such an implicit prior, we construct the posterior as before, and to sample from the posterior we can make use of the \texttt{RegularizedLinearRTO} sampler:
\begin{codebox}
y = Gaussian(A@x, 0.1)
posterior = JointDistribution(x, y)(y=y_data)

samples = RegularizedLinearRTO(posterior).sample(1000).get_samples()
\end{codebox}
Samples from the nonnegativity-constrained posterior are shown in the bottom-right of Figure~\ref{fig:simplest_linear}.

Likewise, if we know that $x_1$ and $x_2$ might be the same, we could use total variation regularization to penalize the difference between the two parameters using:
\begin{codebox}
x = RegularizedGaussian(np.zeros(2), 10,
                        regularization="TV", strength=0.2,
                        geometry=A.domain_geometry)
\end{codebox}
In this code snippet, the strength of the total variation regularization is provided using the \texttt{strength} argument, and a geometry for \texttt{x} needs to be provided, such that a distinction can be made between total variation regularization for one-dimensional signals and two-dimensional images. 

The \texttt{RegularizedLinearRTO} sampler chooses the optimization algorithm based on the chosen constraints and/or regularization. For example, it uses FISTA for nonnegativity constraints and ADMM for total variation. Additional parameters can also be passed to \texttt{RegularizedLinearRTO} to manually tune the optimization algorithms or to choose another optimization algorithm, if applicable. It is also possible to use both constraints and regularization together as easily as providing both \texttt{constraint} and \texttt{regularization} arguments:
\begin{codebox}
x = RegularizedGaussian(np.zeros(2), 10,
                        constraint="nonnegativity",
                        regularization="TV", strength=0.2,
                        geometry=A.domain_geometry)
\end{codebox}

To make RLRTO easy to use, we have implemented various other constraint and regularization options. For constraints we currently provide: \texttt{"nonnegativity"} and \texttt{"box"}  constraints, and for one-dimensional problems we also provide \texttt{"increasing"} , \texttt{"decreasing"} , \texttt{"convex"}  and \texttt{"concave"}. For regularization we currently provide: $l_1$(\texttt{"l1"}) penalty function, also known as LASSO, and anisotropic total variation (\texttt{"TV"}) of the form $\|\lopd{D}\cdot\|_1$ in one and two-dimensions.

Furthermore, for ease of use and to simplify code, two classes named \texttt{ConstrainedGaussian} and \texttt{NonnegativeGaussian} have been added with restricted parameter options. For the popular Gaussian Markov Random Field (GMRF) priors, we also provide the specific classes: \texttt{RegularizedGMRF}, \texttt{ConstrainedGMRF} and \texttt{NonnegativeGMRF}.

\subsubsection{Hyperpriors with RLRTO}
Note that in all the previous examples, we require knowledge of the noise level and need to pick a strength of the prior. To enhance the posterior and improve inference, we can choose to make these hyperparameters unknown and infer them together with the parameter $\vecd{x}$. To achieve this, we put a hyperprior distribution on these hyperparameters. For total variation regularization, such a hierarchical problem could look something like this:
\begin{align*}
    \vecd{x}\,|\,\vecd{y},l,d &= \argmin_{\vecd{x}\in \mathbb{R}_{\geq 0}^n} \left\{\frac{l}{2}\|\lopd{A}\vecd{x}-\hat{\vecd{y}}\|^2_{\lopd{\Sigma}_l^{-1}} + d\|\lopd{D}\vecd{x}\|_1\right\},\\
    l &\sim \Gamma(\alpha_l, \beta_l), \text{ and}\\
    d &\sim \Gamma(\alpha_d, \beta_d),
\end{align*}
where $\lopd{\Sigma}_l^{-1} = l^{-1}\lopd{I}$ and $\Gamma(\alpha, \beta)$ denotes a Gamma distribution with shape $\alpha$ and rate $\beta$.

\begin{figure}
    \centering
    \begin{subfigure}{0.33\textwidth}
        \centering
        \includegraphics[width=\textwidth]{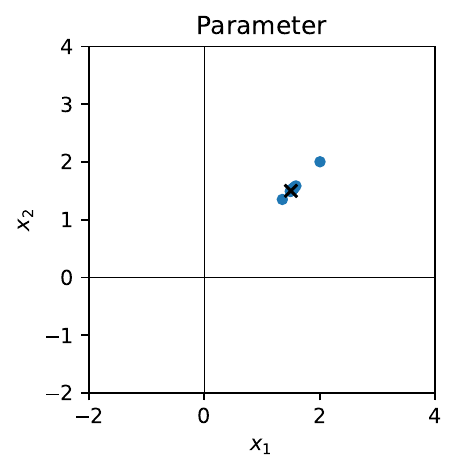}
    \end{subfigure}%
    \begin{subfigure}{0.33\textwidth}
        \centering
        \includegraphics[width=\textwidth]{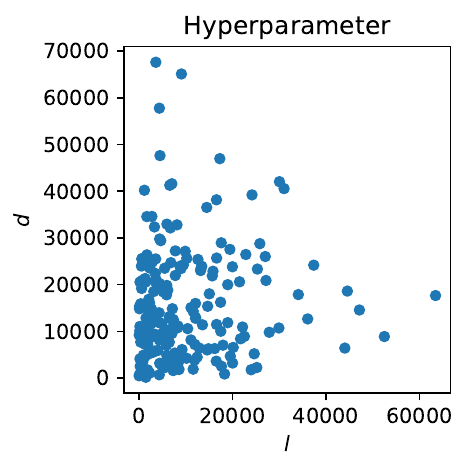}
    \end{subfigure}%
    \caption{Samples of a posterior for a simple linear inverse problem using a regularized Gaussian prior with total variation regularization in a hierarchical model.}
    \label{fig:simplest_linear_hierarchical}
\end{figure}

Note that in the above hierarchical problem there is no prior randomized least squares term $\frac12\|\vecd{x}-\vecd{\hat{\mu}}\|^2_{\lopd{\Sigma}_p^{-1}}$. The corresponding \cuqipy\ code for the model above, together with a Gibbs sampler, is as follows:
\begin{codebox}
d = Gamma(a_d, b_d)
l = Gamma(a_l, b_l)
x = RegularizedUnboundedUniform(regularization="TV",
                                strength=lambda d : d,
                                geometry=domain_geometry)
y = Gaussian(A@x, prec=lambda l : l)
posterior = JointDistribution(d, l, x, y)(y=y_data)

sampling_strategy = {
            'd': Conjugate(),
            'l': Conjugate(),
            'x': RegularizedLinearRTO()}

samples = HybridGibbs(posterior, sampling_strategy).sample(1000).get_samples()
\end{codebox}
Note that the \texttt{RegularizedGaussian} instance is replaced with an instance of \texttt{RegularizedUnboundedUniform}. The \texttt{RegularizedUnboundedUniform} implements the absence of the prior randomized least squares term $\frac12\|\vecd{x}-\vecd{\hat{\mu}}\|^2_{\lopd{\Sigma}_p^{-1}}$ by assigning it a precision matrix $\Sigma_p^{-1}$ of zero, which can be interpreted as an unbounded uniform distribution. 
Samples from this posterior are shown in Figure~\ref{fig:simplest_linear_hierarchical}. Currently, \cuqipy\,supports hierarchical formualtions with RLRTO for all implemented constraint and regularization options, though hierarchical formulations using these regularized distributions has quite some limitations. For details on hierarchical formulations of inverse problems using RLRTO and additional features not mentioned here, see \cite{everink2024uncertainty}.

\subsubsection{User-defined constraints and regularization}
For full flexibility, we also allow users to specify their own constraints and regularization.
For constraints, the user can specify a function that projects any parameters onto their desired constraint set. For example, constraining the parameters to have coefficients larger than $-1$ can be achieved with:
\begin{codebox}
x = ConstrainedGaussian(np.zeros(2), 10,
                        projector=lambda z : np.maximum(z, -1))
\end{codebox}
The projector provided will be used in the underlying solver, in this case FISTA, to sample efficiently.

For user-defined regularization there are two approaches. If the user has an implementation, say \texttt{P}, of the proximal operator for their desired regularization function, then it can be passed as a parameter as follows:
\begin{codebox}
x = RegularizedGaussian(np.zeros(2), 10, proximal=lambda z, s : P(z, s))
\end{codebox}

For regularization taking a more general form like:
\begin{equation*}
    R(\vecd{x}) = f_1(\lopd{L}_1\vecd{x}) + f_2(\lopd{L}_2\vecd{x}) + \ldots,
\end{equation*}
ADMM will be used to sample efficiently and therefore we require the user to provide the proximal operators of the functions $f_i$ paired with the linear operators $L_i$ as a list like $[(\vecd{z}, s \mapsto \text{prox}_{\sigma f_i}(\vecd{z}), \lopd{L}_i)]_{i = 1,2,\dots}$.
For example, consider nonnegativity constrained total variation regularization of the form:
\begin{equation}\label{eq:nnTV}
    R(\vecd{x}) = \|\lopd{D}\vecd{x}\|_1 + \chi_{\mathbb{R}^n_{\geq 0}}(\vecd{x}).
\end{equation}
Denoting by \texttt{P} the proximal operator of the $l_1$ norm, then \eqref{eq:nnTV} can be realized using:
\begin{codebox}
proximal_list = [(lambda z, s : P(z, s), D),(lambda z, _ : np.maximum(z, 0), np.eye(n))]
x = RegularizedGaussian(np.zeros(2), 10, proximal=proximal_list)
\end{codebox}
Here, the first tuple in \texttt{proximal\textunderscore list} corresponds to the proximal operator of the $l_1$ norm and the finite-difference matrix $\lopd{D}$, hence together they represent the regularization term $\|\lopd{D}\vecd{x}\|_1$ in \eqref{eq:nnTV}. The second tuple in \texttt{proximal\textunderscore list} corresponds to the proximal operator of the indicator function of $\chi_{\mathbb{R}^n_{\geq 0}}$, i.e., the Euclidean projection onto $\mathbb{R}^n_{\geq 0}$, and an identity matrix $\lopd{I}$, hence together they represent the regularization term $\chi_{\mathbb{R}^n_{\geq 0}}(\vecd{x})$ in \eqref{eq:nnTV}.

This is also the same way the pre-defined nonnegativity constrained total variation regularization is implemented. However, user-specified constraints and regularization cannot be used for sampling from hierarchical formulations, unless the user specifies their own conjugacy rules.

\subsection{Langevin algorithms}\label{subsec:langevin_in_cuqipy}
This example demonstrates the use of Langevin algorithms with implicit priors in \cuqipy. We consider the following nonlinear variation of the ``probably the simplest Bayesian inverse problem in the world'' from the previous subsection. The inverse problem is written as
\begin{equation}\label{eq:simplest-nonlinear-problem}
    y = x_1^2 + x_2^2 + e, \text{ with } e \sim \mathcal{N}(0, 0.1^2).
\end{equation}
With observed measurement $y_{\text{obs}}=5.176$, the goal is to recover the values of $x_1$ and $x_2$, clearly an underdetermined problem.

\subsubsection{Nonlinear forward model}
In \cuqipy, a nonlinear forward model can be defined using the \texttt{Model} class by providing the routines to compute the \texttt{forward} and (optionally) \texttt{jacobian}, together with \texttt{domain\_geometry} (e.g., 2 for $\mathbb{R}^2$) and \texttt{range\_geometry} (e.g., 1 for $\mathbb{R}$). Note here that we do not specify a geometry object; we pass the number of inputs and outputs instead, which \texttt{Model} uses to create default geometry objects. The forward model for \eqref{eq:simplest-nonlinear-problem} is implemented as follows:
\begin{codebox}
def forward(x):
    return x[0]**2 + x[1]**2
    
def jacobian(x):
    return np.array([[2*x[0], 2*x[1]]])
    
A = Model(forward,
          jacobian=jacobian,
          domain_geometry=2,
          range_geometry=1)
\end{codebox}

Unlike the \texttt{LinearModel} used in \eqref{subsec:cp_explicit}, which infers geometry from the provided matrix, the \texttt{Model} class requires explicit geometry specification due to its reliance on user-defined routines. The Jacobian facilitates gradient-based sampling methods, e.g., ULA \cite{durmus2017nonasymptotic} and NUTS \cite{hoffman2014no}.

In the following, we explore the posterior under different implicit priors, namely unbounded uniform, nonnegativity, $l_1$, and TV.

\subsubsection{Uniform prior}
We start with an unbounded uniform prior over $\mathbb{R}^2$, making the posterior distribution solely determined by the Gaussian likelihood. The likelihood, centered on the circle $x_1^2+x_2^2=5.176$, induces a ring-shaped posterior, reflecting the non-uniqueness of the inverse problem. The \cuqipy{} implementation is:
\begin{codebox}
x = UnboundedUniform(2)
y = Gaussian(A(x), 0.1**2)
posterior = JointDistribution(x, y)(y=y_obs)
\end{codebox}
We use the NUTS sampler to draw samples from this posterior. As shown in Figure~\ref{fig:simplest-nonlinear-likelihood-only}, the samples are concentrated around a circular ring, consistent with the likelihood's geometry.
\subsubsection{Nonnegativity}
Here we choose to impose nonnegativty via an implicit prior. To use implicit priors with the Langevin algorithms in \cuqipy, the user first has to create a \texttt{RestorationPrior} object that is defined with a user-supplied function to perform the restoration (denoising) operation. This object corresponds to \eqref{eq:proximal}. Here we start with the nonnegativity prior, and the restoration operation is defined as a Euclidean projection to the nonnegative orthant:
\begin{codebox}
def nonnegativity_restore(x, restoration_strength):
    return np.maximum(x, 0), None

nn_prior = RestorationPrior(
    nonnegativity_restore,
    geometry=A.domain_geometry
)
\end{codebox}
The user-provided function returns a 2-tuple, with \texttt{maximum(x,0)} being the restored/denoised result and \texttt{None} being a placeholder for potential diagnostic information, as required by the \texttt{RestorationPrior} interface.

Then we need to define a \texttt{MoreauYoshidaPrior} object that corresponds to \eqref{eq:moreau_envelope} with a scalar representing the smoothing strength.

\begin{codebox}
x = MoreauYoshidaPrior(prior=restorator, smoothing_strength = 1e-3)
\end{codebox}
With this \texttt{MoreauYoshidaPrior} object created, we can create the posterior as before and draw samples from it with the ULA sampler:
\begin{codebox}
y = Gaussian(A(x), 0.1**2)
posterior = JointDistribution(x, y)(y = y_obs)
samples = ULA(posterior, scale=1e-3,
              initial_point=np.array([0.5, 0.5])).sample(100000).get_samples()
\end{codebox}

As shown in Figure~\ref{fig:simplest-nonlinear-nonnegativity}, the samples concentrate in the nonnegative quadrant around the arc $x_1^2+x_2^2=5.176$, with some samples deviating slightly into negative regions due to the soft nonnegativity constraint in the Langevin algorithm, in contrast to RLRTO’s hard constraints.

\subsubsection{Sparsity by $l_1$ prior}
To promote sparsity, we use an $l_1$ prior via a \texttt{RestorationPrior} with a soft-thresholding operation, encouraging solutions that lie near the axes. The implementation in \cuqipy{} is as follows:
\begin{codebox}
def l1_restore(x, restoration_strength):
    gamma = 10 * restoration_strength
    return np.multiply(np.sign(x), np.maximum(np.abs(x) - gamma, 0)), None
\end{codebox}

The creation of its corresponding \texttt{MoreauYoshidaPrior} object is the same as demonstrated earlier with the nonnegativity prior and is omitted here and also in the subsequent example to avoid redundancy.

The posterior is multimodal due to the likelihood's circular ring ($x_1^2+x_2^2=5.176$) and the $l_1$ prior's preference for sparse solutions (e.g., near the axes). For this specific problem, the posterior is known to exhibit four distinct modes. In theory, a single long Markov chain initialized at a random point could eventually explore all four modes of the posterior; in practice, this is often inefficient and unreliable for multimodal distributions. Transitions between well-separated modes can be extremely rare, particularly when modes are narrow or isolated, as is the case here. To ensure sufficient exploration and avoid poor mixing, we run four independent chains, each initialized near a different mode. As seen in Figure~\ref{fig:simplest-nonlinear-l1}, the samples cluster around the intersections of the axes and the circular ring, reflecting the multimodal nature of the posterior and the sparsity-encouraging behaviour as intended.

\subsubsection{TV prior}\label{sec:tv_prior_cuqipy}
\begin{figure}[tb]
\centering
\subfloat[][Uniform]{\includegraphics[width=0.24\linewidth]{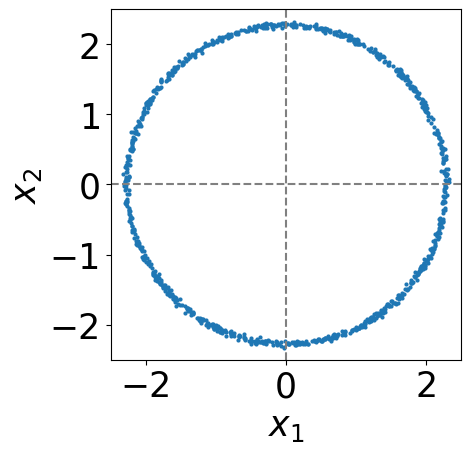}\label{fig:simplest-nonlinear-likelihood-only}}\hfill
\subfloat[][Nonnegativity]{\includegraphics[width=0.24\linewidth]{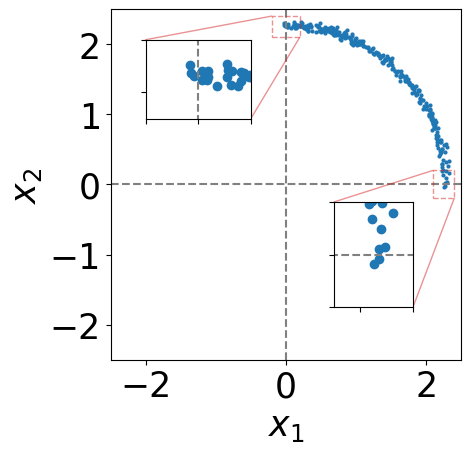}\label{fig:simplest-nonlinear-nonnegativity}}\hfill
\subfloat[][$l_1$]{\includegraphics[width=0.24\linewidth]{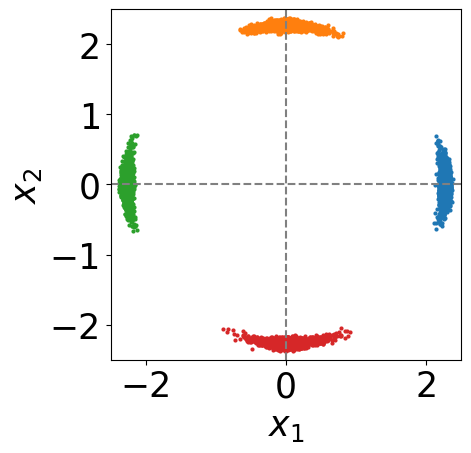}\label{fig:simplest-nonlinear-l1}}\hfill
\subfloat[][TV]{\includegraphics[width=0.24\linewidth]{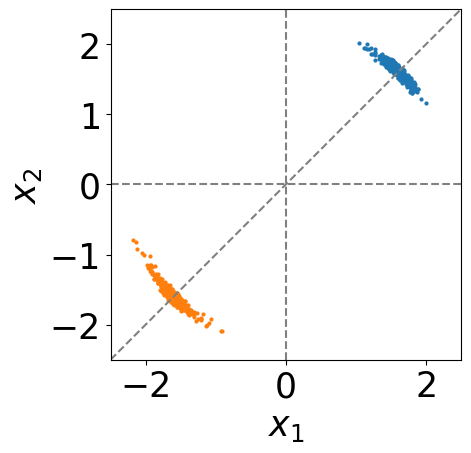}\label{fig:simplest-nonlinear-tv}}

\caption{Posterior samples from solving the simple nonlinear inverse problem using different priors with the Langevin algorithm. Colors denote distinct Markov chains that are used to capture multimodal posteriors with $l_1$ and TV priors.}
\label{fig:simplest-nonlinear}
\end{figure}

This experiment uses a TV prior to promote solutions where $x_1\approx x_2$, reflecting the prior's preference of minimal component differences. Here we create a \texttt{RestorationPrior} object with the help of \texttt{denoise\_tv\_chambolle} from \codefont{scikit-image} \cite{scikit-image} and the TV weight is set to be $5$:
\begin{codebox}
def tv_restore(x, restoration_strength):
    return denoise_tv_chambolle(x, weight = 5 * restoration_strength,
                                max_num_iter = 100), None
\end{codebox}
The posterior is again multimodal due to the effect of the TV prior on the circular likelihood, and therefore, we use two Markov chains from different starting points. As shown in Figure~\ref{fig:simplest-nonlinear-tv}, the samples are more concentrated around the intersections between the diagonal line ($x_1=x_2$) and the circular ring, reflecting the balance between the prior and the likelihood.

\subsubsection{PnP prior}
We emphasize that, to define an implicit prior for Langevin algorithms in \cuqipy, the user must provide two objects: a \texttt{RestorationPrior}, which implements a restoration operation, and an object that provides the approximated gradient. As seen before, this object can be a \texttt{MoreauYoshidaPrior}, aligning with the intuition of MYULA. In the case of an MMSE denoisers, we provide \texttt{TweediePrior} as an alias of \texttt{MoreauYoshidaPrior} to align with the formulation of PnP-ULA. In this section, we demonstrate the general setup of PnP prior in \cuqipy, without applying it to a particular problem.



With PnP-ULA, a \texttt{RestorationPrior} could be defined just as before, with \texttt{blackbox\_denoise} being a black box function that performs restoration on a given set of parameters. The implementation is as follows:
\begin{codebox}
def pnp_restore(x, restoration_strength):
    return blackbox_denoise(x, restoration_strength), None

pnp_prior = RestorationPrior(
    pnp_restore,
    geometry=A.domain_geometry
)
\end{codebox}
This \texttt{RestorationPrior} is then incorporated into a \texttt{TweediePrior} object as shown below:
\begin{codebox}
x = TweediePrior(prior=pnp_prior, smoothing_strength = smoothing_strength)
\end{codebox}
In this context, \texttt{smoothing\_strength} represents the noise variance used to train an MMSE denoiser.

\section{Application case studies}\label{sec:examples}
We provide various application cases studies that demonstrate the application of implicit priors across different commonly considered inverse problems.
All Python code and notebooks used in the following examples are available at \href{https://doi.org/10.5281/zenodo.17115255}{https://doi.org/10.5281/zenodo.17115255}.

\subsection{Deblurring with constraints}
\label{subsection:deblurring}

Consider solving a continuous deconvolution problem:
\begin{equation}\label{eq:continuous_deconvolution}
    y(t) = \int_{0}^{1} x(t-s)\phi_{\sigma}(s) \text{d}s,\quad \text{ with }\quad \phi_{\sigma}(s) := \frac{1}{\sigma\sqrt{2\pi}}\exp\left(-\frac{1}{2}\left(\frac{s}{\sigma}\right)^2\right),
\end{equation}
and $\sigma = 10$.

\begin{figure}
\centering
\begin{subfigure}{.9\textwidth}
  \centering
  \includegraphics[width=\linewidth]{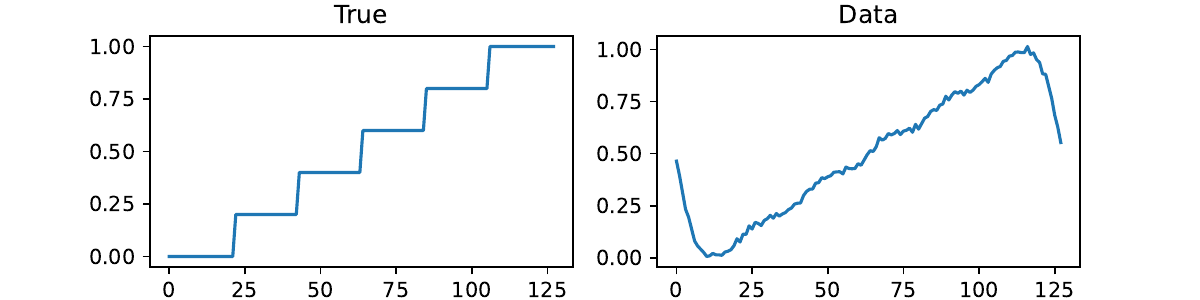}
  \caption{Deconvolution ground truth observed data}
  \label{fig:stair_data}
\end{subfigure}
\begin{subfigure}{.9\textwidth}
  \centering
  \includegraphics[width=\linewidth]{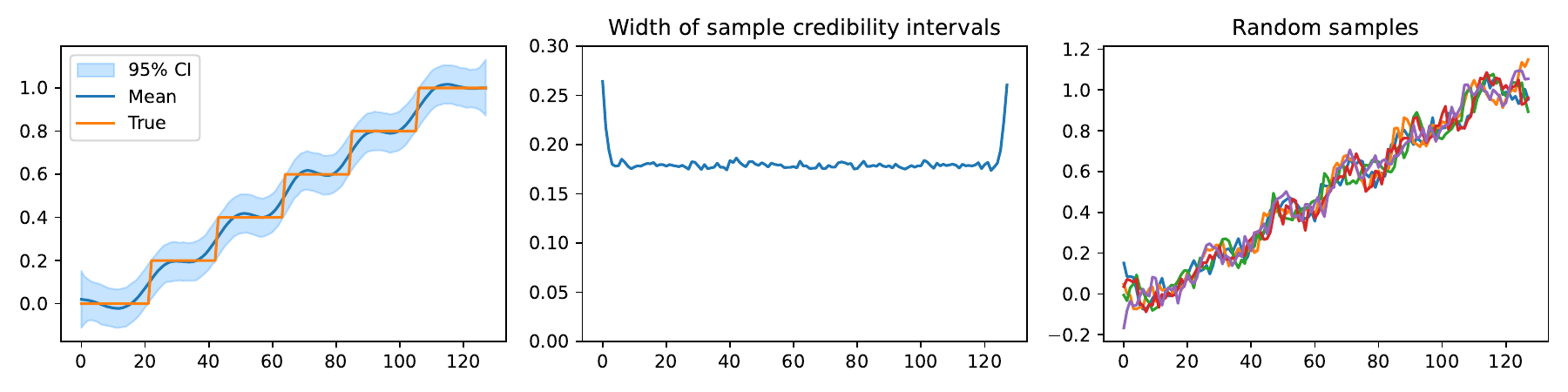}
  \caption{Deconvolution results without constraints.}
  \label{fig:stair_results_gmrf}
  
  \includegraphics[width=\linewidth]{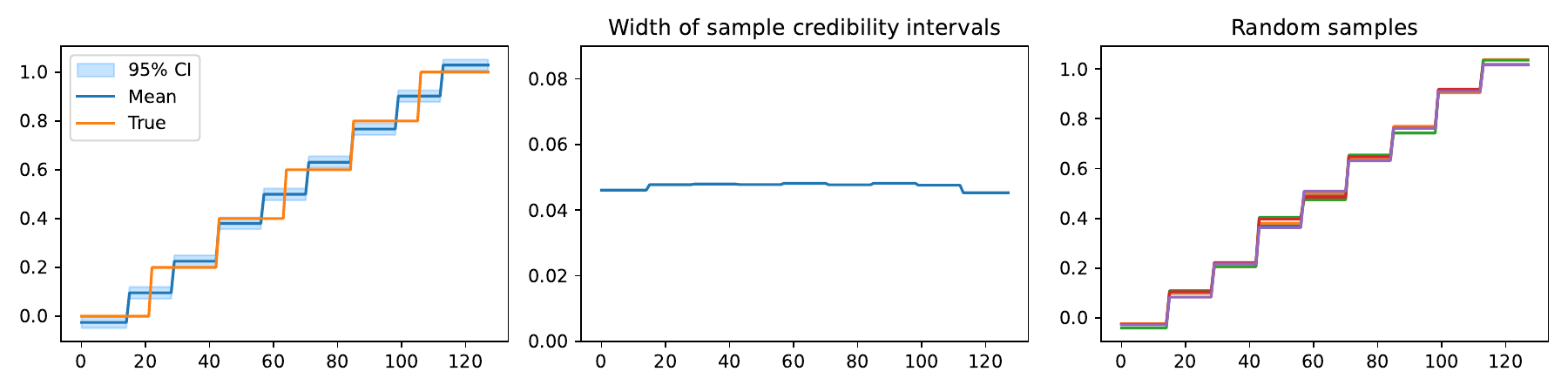}
  \caption{Deconvolution results with a step expansion.}
  \label{fig:stair_results_stepexpansion}
  
  \includegraphics[width=\linewidth]{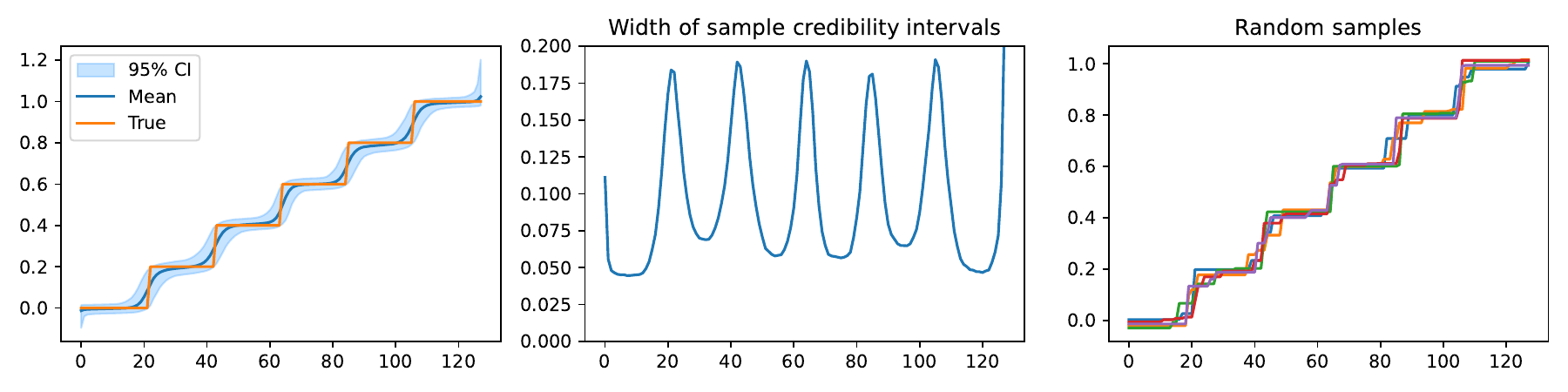}
  \caption{Deconvolution results with monotonicity constraints.}
\label{fig:stair_results_monotone}
    
\end{subfigure}
\caption{One-dimensional deconvolution problem of a monotone/staircase signal. Note that the samples of the constraint priors in \ref{fig:stair_results_stepexpansion} and \ref{fig:stair_results_monotone} show the same staircase behaviour as the underlying truth.}
\label{fig:deconvolution}
\end{figure}

Assuming we know that the signal $x(t)$ is defined on an interval $[0, 1]$, consists of finitely many piecewise constant regions and is increasing. We discretize $x(t)$ using $n = 128$ equidistant points and discretize the integral in \eqref{eq:continuous_deconvolution} using a midpoint rule resulting in $m = 128$ measurements. Including noise, we write the finite dimensional problem as:
\begin{equation*}
    \vecd{y} = \lopd{A}\vecd{x} + \vecd{e},
\end{equation*}
where $\vecd{x}, \vecd{y}, \vecd{e} \in \mathbb{R}^{128}$, $\lopd{A} \in \mathbb{R}^{128\times 128}$ and $\vecd{e}\sim\mathcal{N}(\vecd{0}, 10^{-3}\lopd{I})$. The used ground truth and observed measurements are shown in Figure~\ref{fig:stair_data}.

To show the versatility of priors in \cuqipy, we consider three different ones, one unconstrained and two constrained:
\begin{itemize}
    \item First, for reference we consider a Gaussian Markov random field (GMRF) \cite{rue2005gaussian} with Neumann boundary conditions and precision $500$, i.e., we assume that $x_{i+1} - x_i \sim \mathcal{N}(0, 500^{-1})$. This is an explicit prior, and due to the boundary conditions it is also improper. Results, including mean, credible intervals, credible widths and random samples, are shown in Figure~\ref{fig:stair_results_gmrf}.
    \item Next, we assume that the signal is piecewise constant. We consider $k = 9$ piecewise constant components and a transformation $\theta: \mathbb{R}^k \rightarrow \mathbb{R}^n$ that the $k$ component values in $\vecd{z}$ to a full $n$ long discretized signal $\vecd{x}$. We furthermore assume that $\vecd{z} \sim \mathcal{N}(\vecd{0}, \lopd{I})$. As we will sample on $\vecd{z}$, we consider it a proper, implicit prior. Results with this prior are shown in Figure~\ref{fig:stair_results_stepexpansion}.
    \item Finally, we consider the RLRTO approach. For this, we start out with assuming a Gaussian prior with zero mean and covariance $0.1\lopd{I}$. For constraint, we will assume the parameter $\vecd{x}$ is monotonic, in particular that it is non-decreasing. This is an improper, implicit prior. Results with this prior are shown in Figure~\ref{fig:stair_results_monotone}.
\end{itemize}

Comparing the three different priors, we see in Figure~\ref{fig:deconvolution} that incorporating the constraints makes the samples look a lot more like the ground truth, distinctive from the very noisy-looking unconstrained samples. However, incorrectly specifying the constraints can make the inference overly confident, as is the case with specifying the wrong number of steps. Whilst widths of the credible intervals for the step expansion are still relatively equal over the whole signal, the credible interval widths for the monotonicity constraints peak at the jumps in the signal, which aligns with over smoothing of the mean at these jumps. 

\subsection{Various inverse problems with the Poisson equation}
\label{subsection:poisson}

This subsection illustrates the use of implicit priors in addressing inverse problems governed by the Poisson equation. Widely used in fields like fluid dynamics, chemical diffusion, and electrostatics, the Poisson equation allows us to showcase the applicability of the implicit prior framework. We explore the inference of the source term, boundary condition, and conductivity from noisy measurements of the potential, each considered separately with tailored priors and methods, i.e., RLRTO for inference of the source term and boundary condition and MYULA for the conductivity.


In the following, we outline the common settings for all of the experiments conducted in this subsection, followed by detailed configurations for each case. 

\subsubsection{Common settings}
\label{subsection:poisson-common-settings}
The Poisson equation is as follows
\begin{equation}
\begin{aligned}
    \nabla\cdot(\kappa \nabla u) &= f&& \text{in $\Omega \subset \mathbb{R}^d$}, \\
    u &= g&& \text{on $\partial\Omega$},
\end{aligned}
    \label{eq:poisson}
\end{equation}
where $\kappa$ is the conductivity, $u$ is the potential, $f$ is the source term, $g$ is the Dirichlet boundary value and $d$ is the physical domain dimension, $d=1,2,\;\text{or}\; 3$. The finite element method (FEM) is employed to solve \eqref{eq:poisson}, where the domain $\Omega$ is partitioned into non-overlapping elements. Conductivity, source term, boundary value, and potential are potentially spatially varying; they are all discretized with piecewise linear basis functions and we denote the discretized version of them by $\vecd{\kappa}$, $\vecd{f}$, $\vecd{g}$ and $\vecd{u}$, respectively.

Following \eqref{eq:linear_additive_model}, we can write the inverse problem for $\vecd{\phi}\in \{\vecd{f}, \vecd{g}, \vecd{\kappa}\}$ as
\begin{equation}
    \vecd{y} = \vecd{u} + \vecd{e} = \lopd{A}_{\vecd{\phi}}(\vecd{\phi}) + \vecd{e},
\end{equation}
where $\lopd{A}_{\vecd{\phi}}$ denotes the forward operator solving \eqref{eq:poisson} for $\vecd{u}$ with $\vecd{\phi}$, and the goal of the inverse problem is to infer $\vecd{\phi}$ from $\vecd{y}$. Each of the inverse problems is ill-posed because multiple fields of $\vecd{\phi}$ may yield similar potential measurements and because small noise perturbations can significantly distort the reconstruction, and thus, proper prior information is needed to regularize it.

In each experiment of this subsection, for a given ground truth of $\vecd{\phi}_{\text{true}}$, we apply $\lopd{A}_{\vecd{\phi}}$ to get the exact data of $\vecd{u}_{\text{exact}}$. Measurements $\vecd{y}_\text{obs}$ are taken as the values of  $\vecd{u}$ at every discretization node of the mesh, and we assume additive Gaussian noise, so $\vecd{e} \sim \mathcal{N}(\vecd{0},\sigma_e^2\vecd{I})$.

Following standard conventions in the field of partial differential equations, in this subsection, $x$ denotes spatial coordinates, instead of parameters. In the following, inference experiments are based on Poisson problems of varying dimensionality: $\vecd{f}$ is inferred from a one-dimensional problem, while both $\vecd{g}$ and $\vecd{\kappa}$ are inferred from two-dimensional problems.

\subsubsection{Source term $\vecd{f}$}
\label{subsection:poisson-source-term}
In this first experiment with the Poisson equation, we consider inferring the source term $\vecd{f}$ with a constraint of nonnegativity, i.e., $\vecd{f}\geq 0$. We apply RLRTO, which is applicable here since the forward operator $\lopd{A}_{\vecd{f}}$ is linear with respect to $\vecd{f}$, to include a GMRF prior with a nonnegativity constraint. 

Specifically, we consider \eqref{eq:poisson} on a one-dimensional unit interval $\Omega$ partitioned into a uniform mesh of 128 elements. The exact $\vecd{f}_{\text{true}}$ is constructed by truncating a sinusoid function at 0, i.e.,
\begin{equation}
    f_{\text{true}}(x) = \max(\sin(4\pi x), 0),\quad 0\leq x \leq 1.
\end{equation}
The conductivity $\kappa$ is fixed at 1 throughout the domain, and the boundary value $g$ is fixed at 0 at both endpoints. To assess the sensitivity of the method to the magnitude of noise, three sets of synthetic data are generated at various noise levels, i.e., $\sigma_e\in \{0.3\cdot 10^{-3},  10^{-3}, 3\cdot 10^{-3}\}$. Sampling with RLRTO is performed within a hierarchical formulation similar to \eqref{fig:simplest_linear_hierarchical}, with the GMRF precision being inferred simultaneously, and as hyperprio on the precision we put a Gamma distribution $\Gamma(1, 10^{-8})$, which is relatively uninformative.

\begin{figure}[tbp]
    \centering
    \subfloat[\centering $\vecd{y}_\text{obs}$ with $\sigma_e=0.3\cdot 10^{-3}$]{{\includegraphics[width=0.32\linewidth]{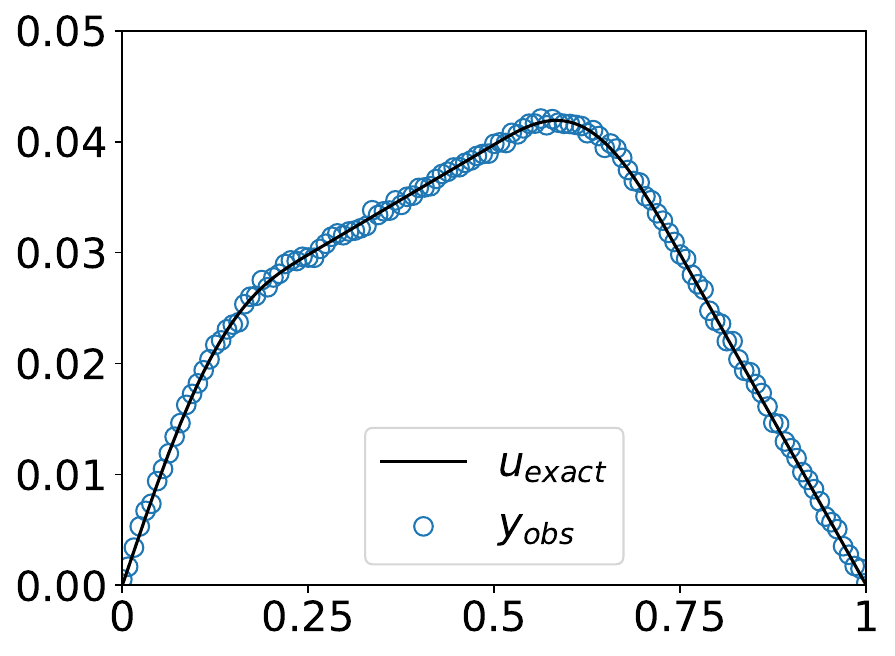} }}
    \hfill
    \subfloat[\centering Nonnegativity]{{\includegraphics[width=0.32\linewidth]{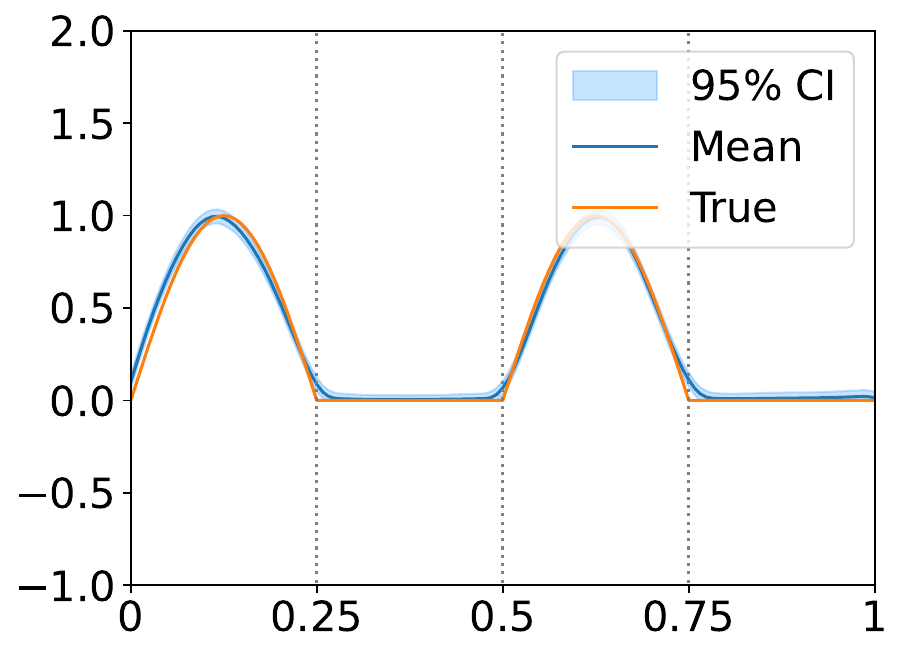} }}%
    \hfill
    \subfloat[\centering GMRF reference]{{\includegraphics[width=0.32\linewidth]{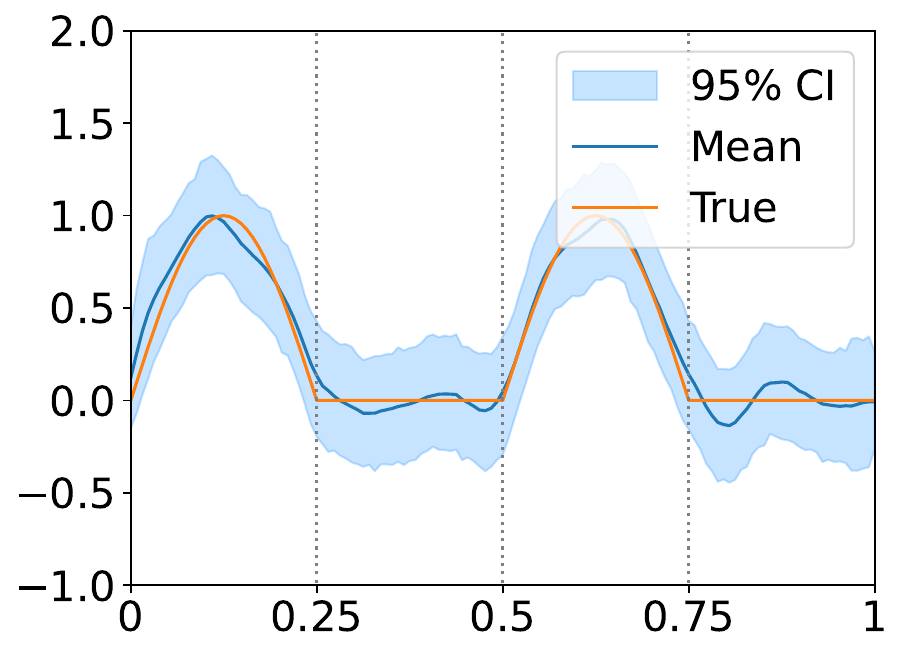} }}\\
    \subfloat[\centering $\vecd{y}_\text{obs}$ with $\sigma_e= 10^{-3}$]{{\includegraphics[width=0.32\linewidth]{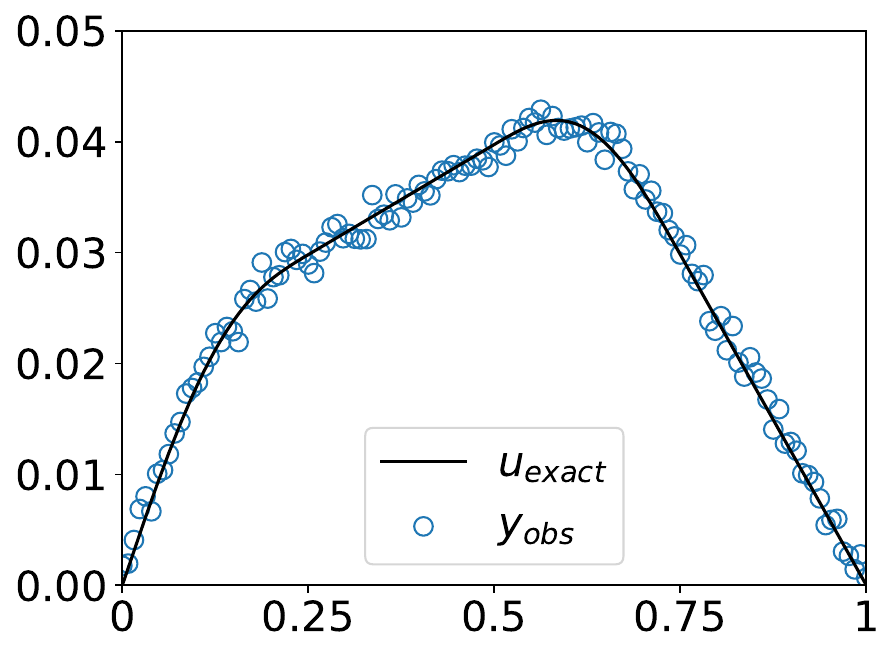} }}
    \hfill
    \subfloat[\centering Nonnegativity]{{\includegraphics[width=0.32\linewidth]{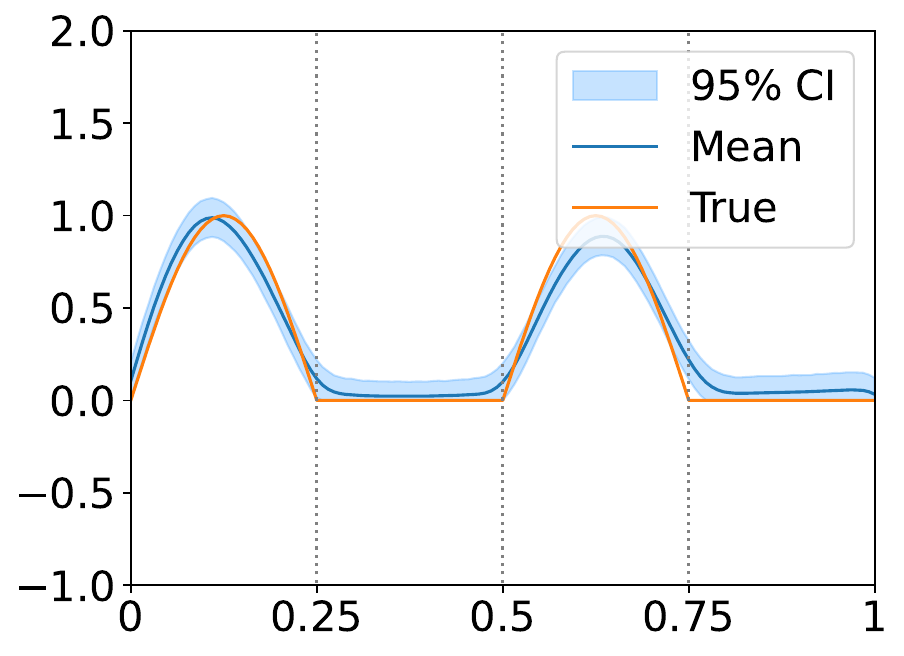} }}%
    \hfill
    \subfloat[\centering GMRF reference]{{\includegraphics[width=0.32\linewidth]{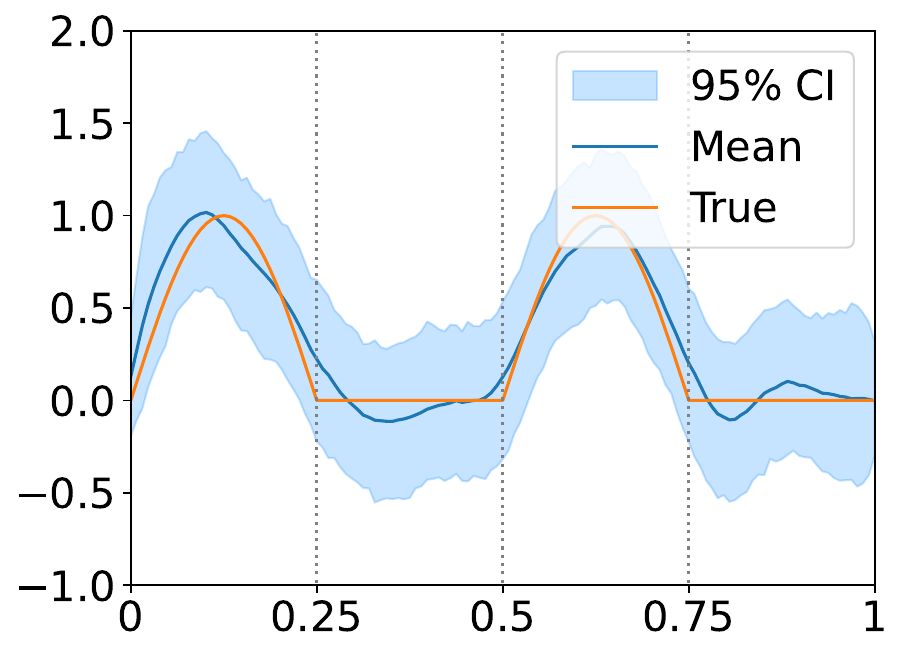} }}\\
    \subfloat[\centering $\vecd{y}_\text{obs}$ with $\sigma_e=3\cdot 10^{-3}$]{{\includegraphics[width=0.32\linewidth]{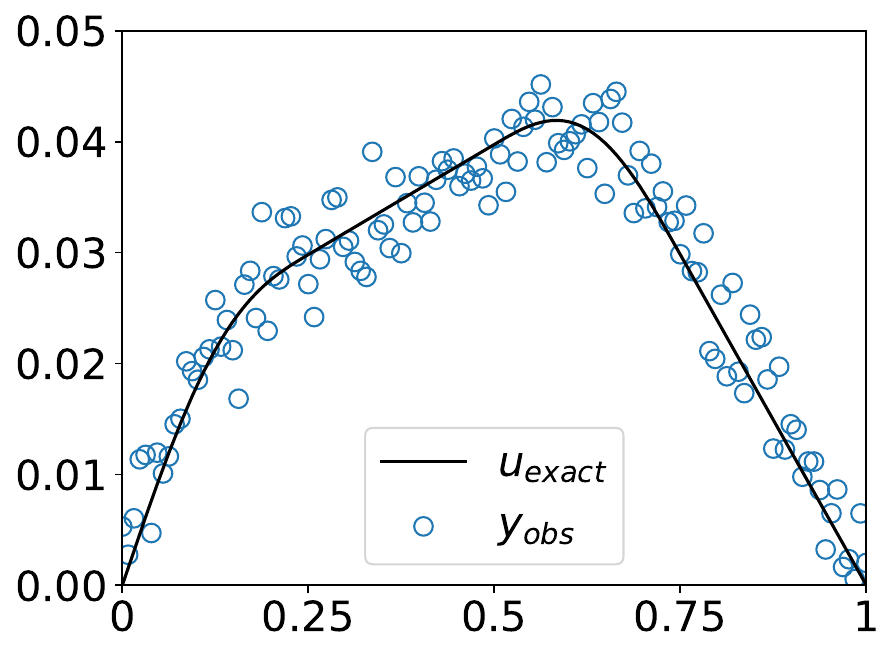} }}
    \hfill
    \subfloat[\centering Nonnegativity]{{\includegraphics[width=0.32\linewidth]{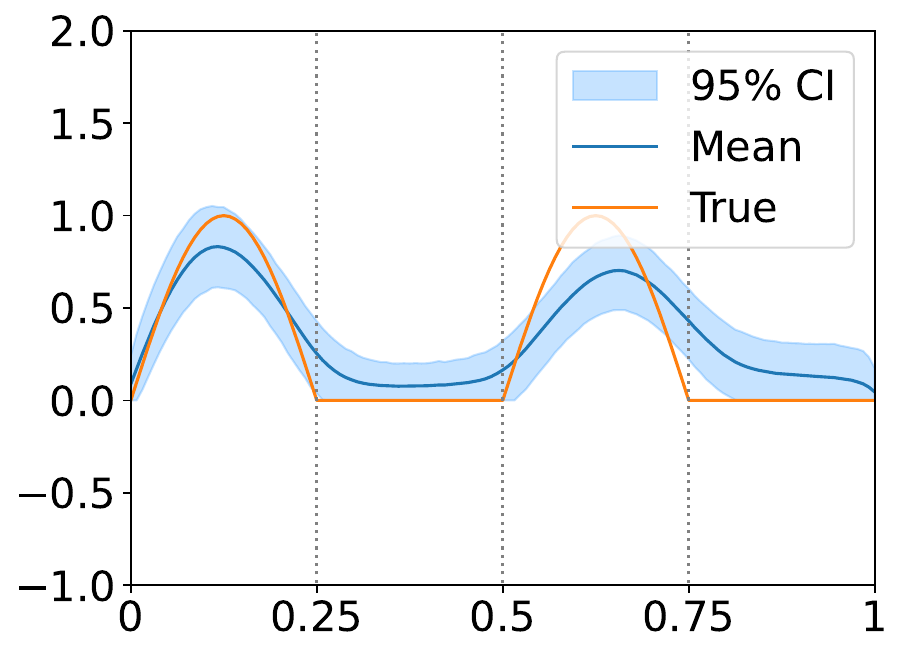} }}%
    \hfill
    \subfloat[\centering GMRF reference]{{\includegraphics[width=0.32\linewidth]{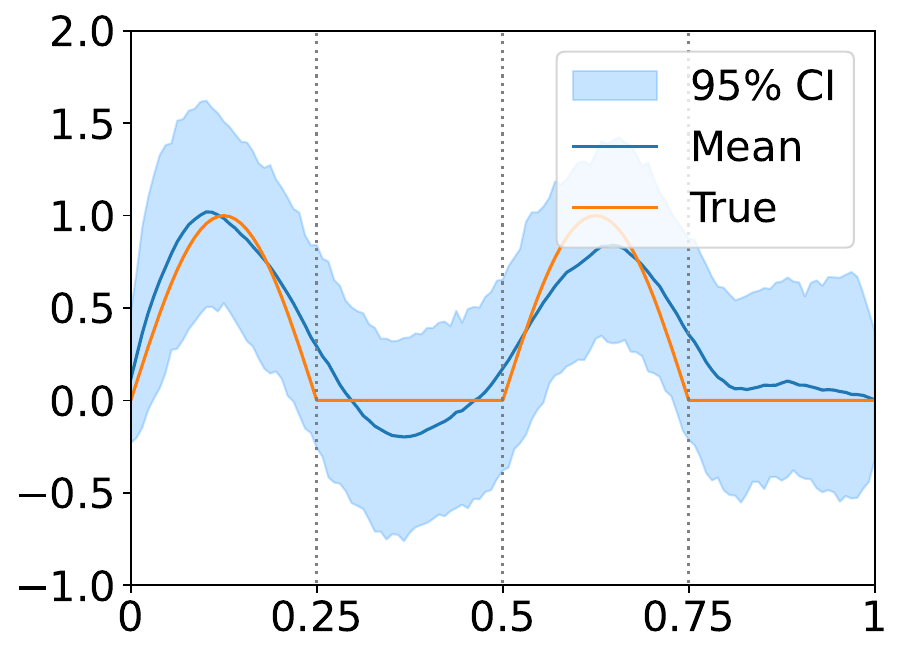} }}
    \caption{Inference of source term $\vecd{f}$ with nonnegativity constraint through RLRTO. The three rows correspond to experiments with noise level $\sigma_e=0.3\cdot 10^{-3}$, $10^{-3}$, and $3\cdot 10^{-3}$, respectively. The left column shows the measurement against the exact data, the middle column shows inference results with RLRTO, and the right column shows reference inference with LRTO. In all cases, the precision of the GMRF prior is inferred together with the source function through a hierarchical formulation.}%
    \label{fig:pde-source}%
\end{figure}

The results are shown in Figure~\ref{fig:pde-source}, which is organized into three rows corresponding to the three sets of data. In each row, the left column shows the measurement data, the middle column shows the posterior statistics (the mean and the 95\% credible interval against the exact $\vecd{f}_{\text{true}}$) from RLRTO, and the right column shows the reference posterior statistics. The reference posterior is obtained by applying a GMRF prior but with no nonnegativity constraint.

Compared with the reference results with GMRF priors only, the nonnegativity constraint applied through RLRTO significantly reduces the posterior variance across all noise levels. This reduction is particularly pronounced in the vicinity of values that are precisely zero, such as those in the interval between 0.25 and 0.5. It is noteworthy that RLRTO strictly enforces the nonnegativity constraint on each posterior sample.

\subsubsection{Boundary value $\vecd{g}$}
\label{subsection:poisson-boundary-value}
This experiment explores recovering the boundary value on the left boundary of a two-dimensional unit square domain, while the boundary value on the rest of the boundary is assumed to be known as 0. Under a slight abuse of notation, we still refer to it as $\vecd{g}$ though it is only a portion of the whole boundary. Similar to the previous experiment, the operator $\lopd{A}_{\vecd{g}}$ is linear in $\vecd{g}$ and thus the RLRTO framework is applicable. Particularly, we consider a GMRF prior with TV regularization, i.e., $\|\lopd{D} \vecd{g}\|_1$ with $\lopd{D}$ being the first-order finite difference operator on the nodes of the left boundary, to promote piecewise-constant features.

\begin{figure}[tb]
    \centering
    \subfloat[\centering $\vecd{y}_\text{obs}$ with $\sigma_e=3\cdot 10^{-1}$]{\includegraphics[width=0.32\linewidth]{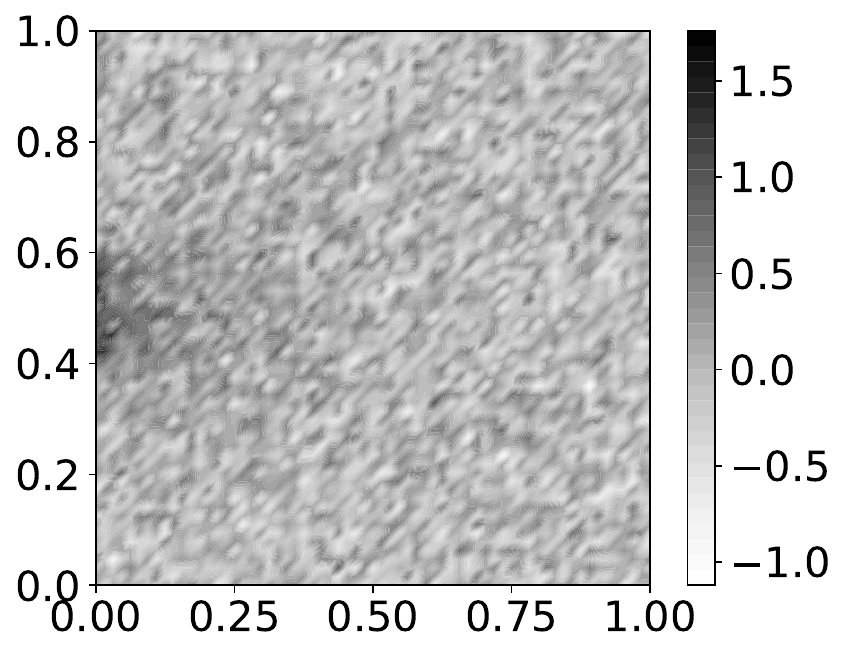}
    \label{fig:pde-boundary-value-data}}
    \hfill
    \subfloat[\centering TV]{\includegraphics[width=0.32\linewidth]{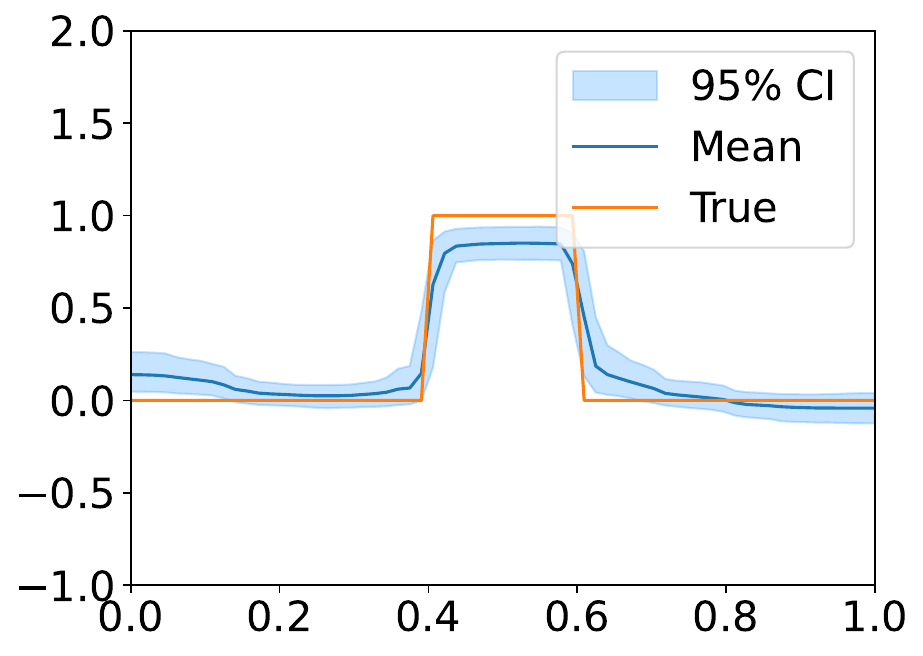}
    \label{fig:pde-boundary-value-tv-post}}%
    \hfill
    \subfloat[\centering TV samples]{\includegraphics[width=0.32\linewidth]{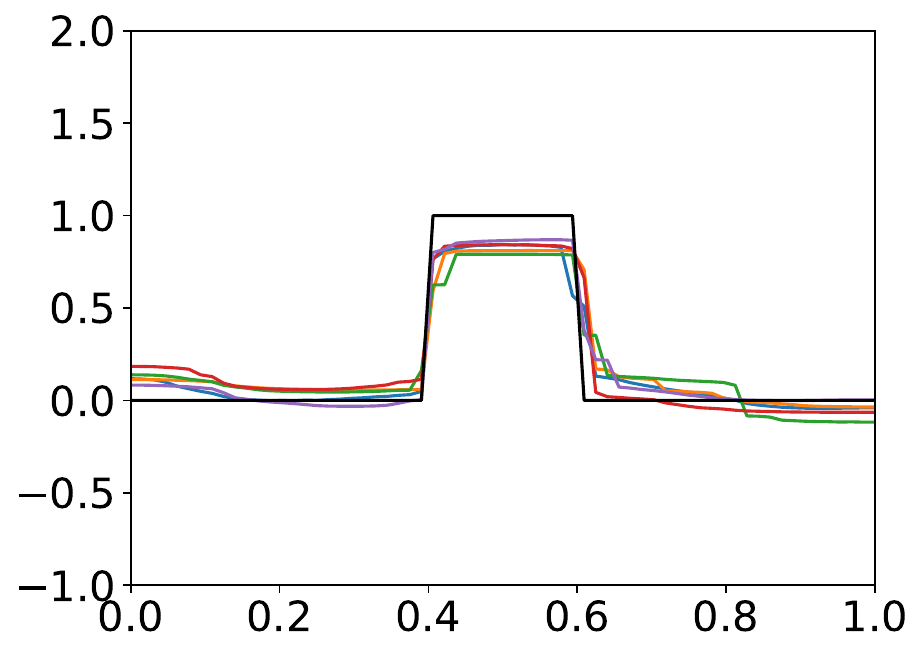}
    \label{fig:pde-boundary-value-tv-samples}}
    \caption{Inference of boundary value $\vecd{g}$ with a GMRF prior and TV regularization through RLRTO.}%
    \label{fig:pde-boundary-value}%
\end{figure}

\begin{figure}[tb]
    \centering
    \subfloat[\centering Reference]{\includegraphics[width=0.32\linewidth]{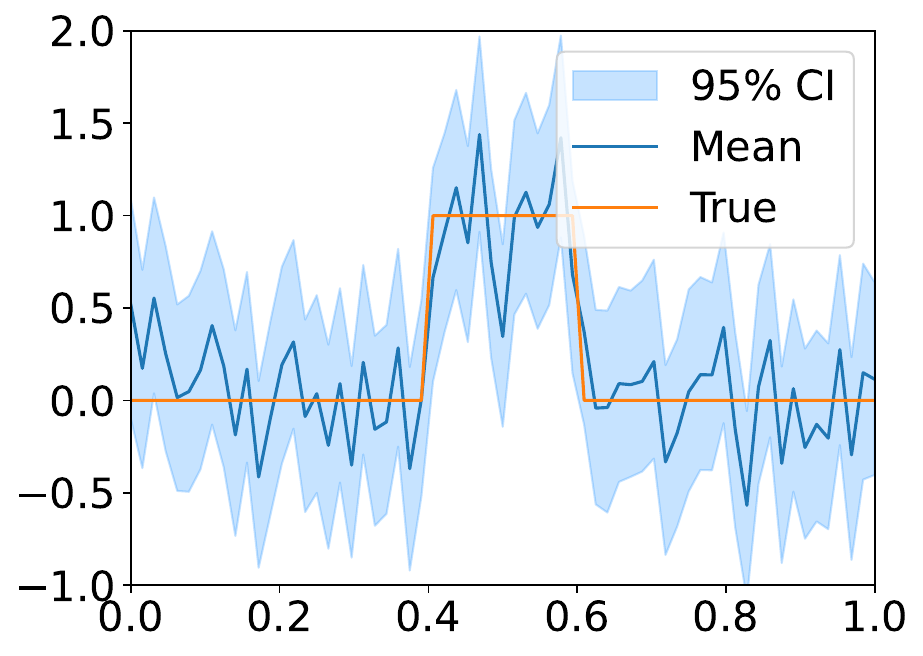}
    \label{fig:pde-boundary-value-reference-post}}
    \subfloat[\centering Reference samples]{\includegraphics[width=0.32\linewidth]{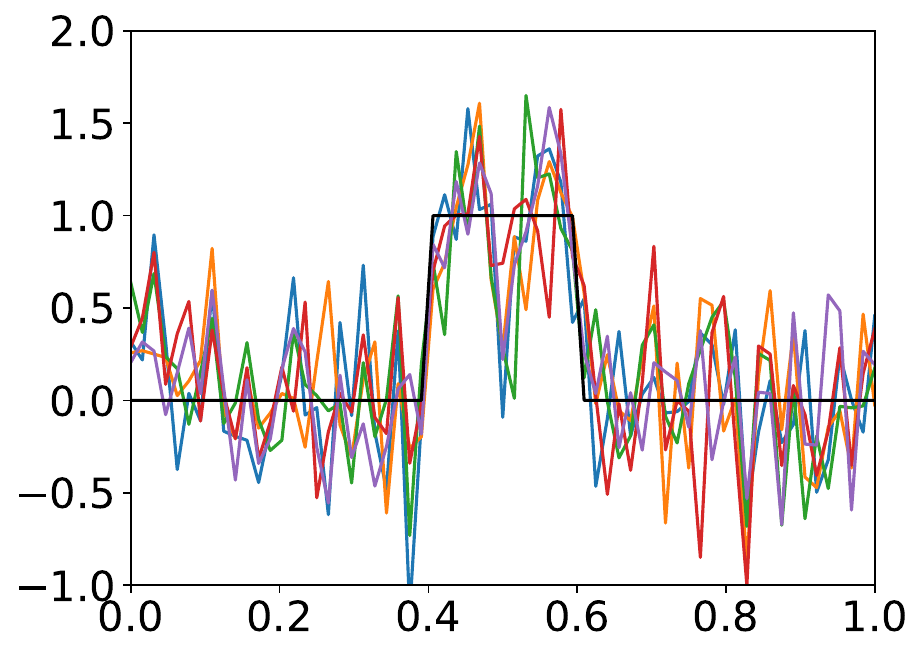}
    \label{fig:pde-boundary-value-reference-samples}}
    \caption{Inference of boundary value $\vecd{g}$ with a GMRF prior only.}%
    \label{fig:pde-boundary-value-reference}%
\end{figure}

The unit square domain is partitioned into a uniform mesh of 64$\times$64$\times$2 triangular elements, yielding $65$ nodes on the left boundary. The conductivity $\kappa$ is fixed at 1, and the source term is set to be 0. The ground truth boundary condition $\vecd{g}_{\text{true}}$ on the left boundary is defined as follows:
\begin{equation}
    g_{\text{true}}(x) = \begin{cases}
        1,\quad &\text{if } 0.4\leq x < 0.6, \text{ and}\\
        0,\quad &\text{otherwise.}
    \end{cases}
\end{equation}
The noise level $\sigma_e$ is set to 0.3 and the generated measurements $\vecd{y}_{\text{obs}}$ is shown in Figure~\ref{fig:pde-boundary-value-data}.

When sampling with RLTRO, the precision of the GMRF is set to 0.1 and the strength of the TV regularization is heuristically set to 25. The posterior based on 1,000 samples is shown in Figure~\ref{fig:pde-boundary-value-tv-post}, and five samples are plotted in Figure~\ref{fig:pde-boundary-value-tv-samples}. As a reference, we also perform the same experiment but with only the GMRF of the same precision, i.e., without TV regularization, and show the results in Figure~\ref{fig:pde-boundary-value-reference}. The results show that RLRTO with TV regularization correctly recovers the general trend of the ground truth, particularly in identifying its jumps, while the reference with a GMRF prior fails to. Oscillations in the samples are significantly reduced by the TV regularization. The RLRTO samples also show a clear staircase effect, which is what we expect from TV regularization.

\subsubsection{Conductivity $\vecd{\kappa}$}
\label{subsection:poisson-conductivity}

This experiment applies MYULA to infer the conductivity field $\vecd{\kappa}$. In contrast to earlier experiments' focus on linear mappings, here we focus on the nonlinear mapping $\lopd{A}_{\vecd{\kappa}}$ from $\vecd{\kappa}$ to $\vecd{u}$, resulting in a non-Gaussian likelihood. Thus, the RLRTO is not applicable in this case, and we resort to the MYULA framework to include a TV prior.

\begin{figure}[tb]
    \centering
    \subfloat[$\vecd{\kappa}_\text{true}$]{\includegraphics[width=0.32\linewidth]{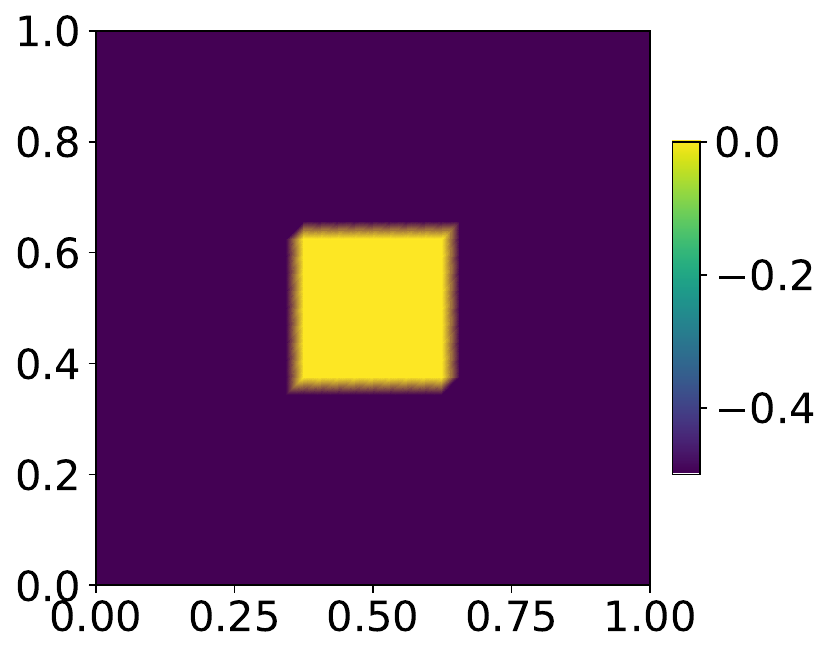}
    \label{fig:pde-my-ula-k-true}}
    \hfill
    \subfloat[$\vecd{u}_{\text{exact}}$]{\includegraphics[width=0.32\linewidth]{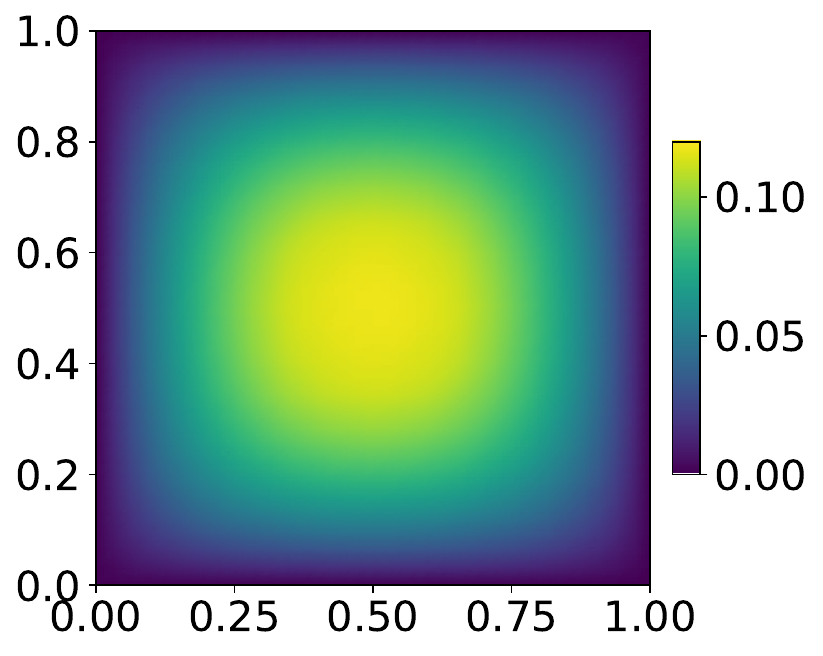} \label{fig:pde-my-ula-u-exact}}%
    \hfill
    \subfloat[$\vecd{y}_\text{obs}$]{\includegraphics[width=0.32\linewidth]{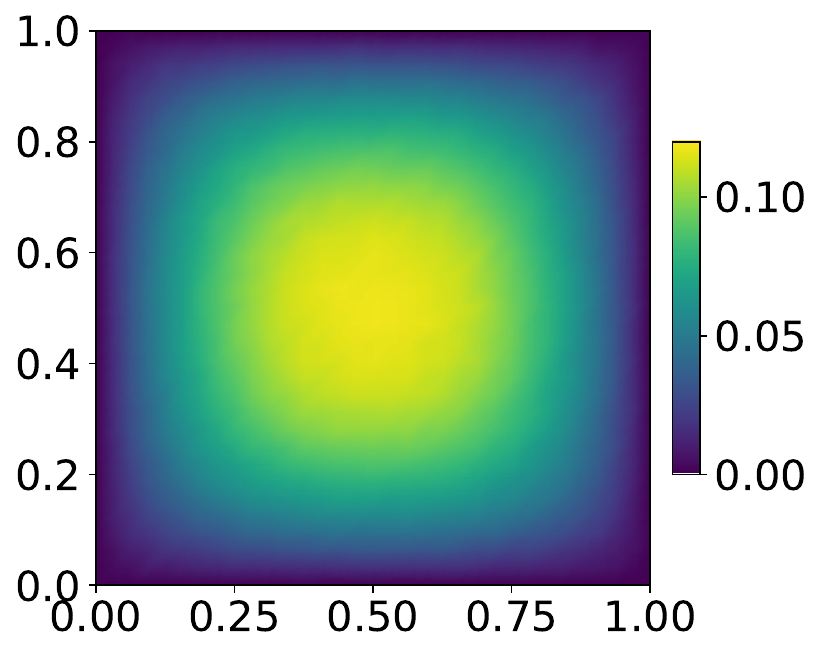} 
    \label{fig:pde-my-ula-y-obs}}
    \\
    \subfloat[Posterior mean]{\includegraphics[width=0.32\linewidth]{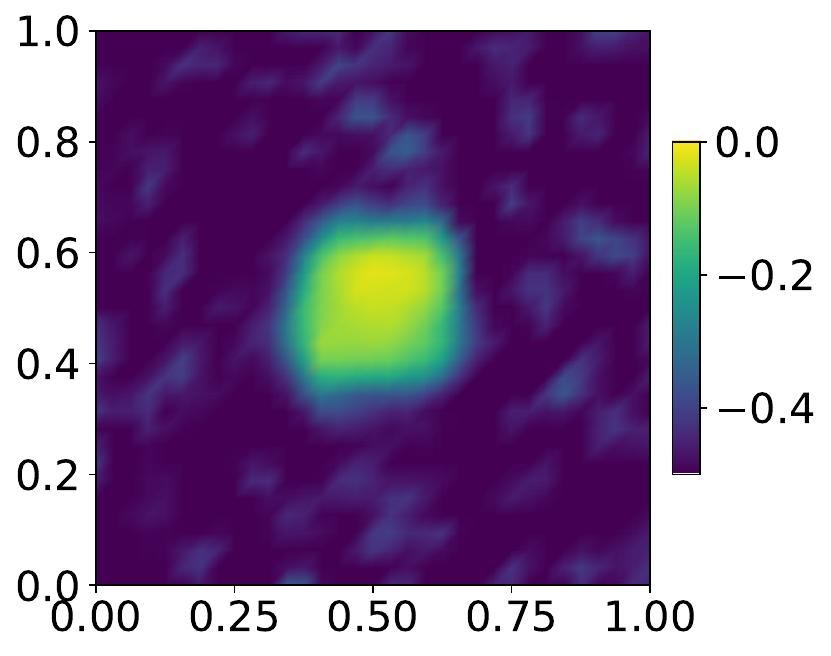}
    \label{fig:pde-my-ula-post-mean}}
    \hfill
    \subfloat[Posterior standard deviation]{\includegraphics[width=0.312\linewidth]{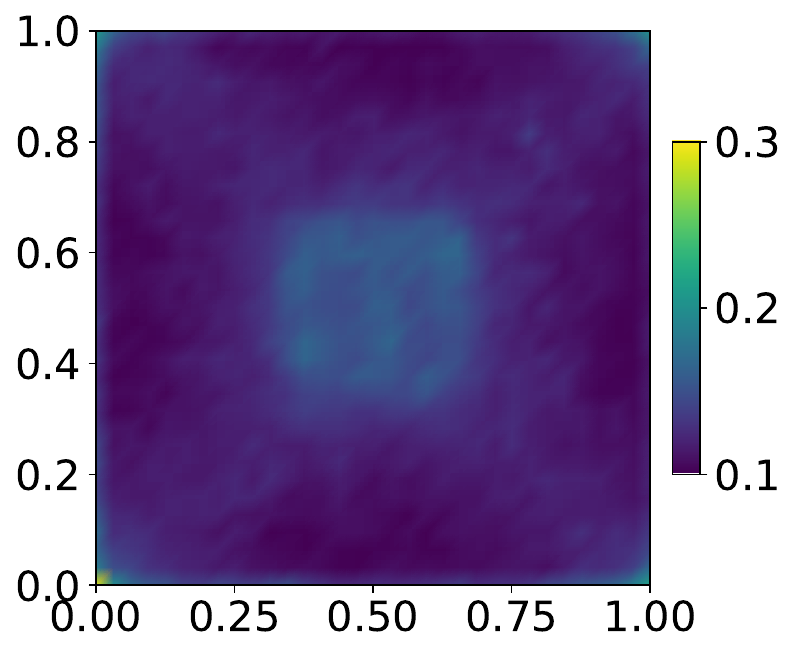} \label{fig:pde-my-ula-post-std}}%
    \hfill
    \subfloat[Posterior at $x_2=0.5$]{\includegraphics[width=0.3\linewidth]{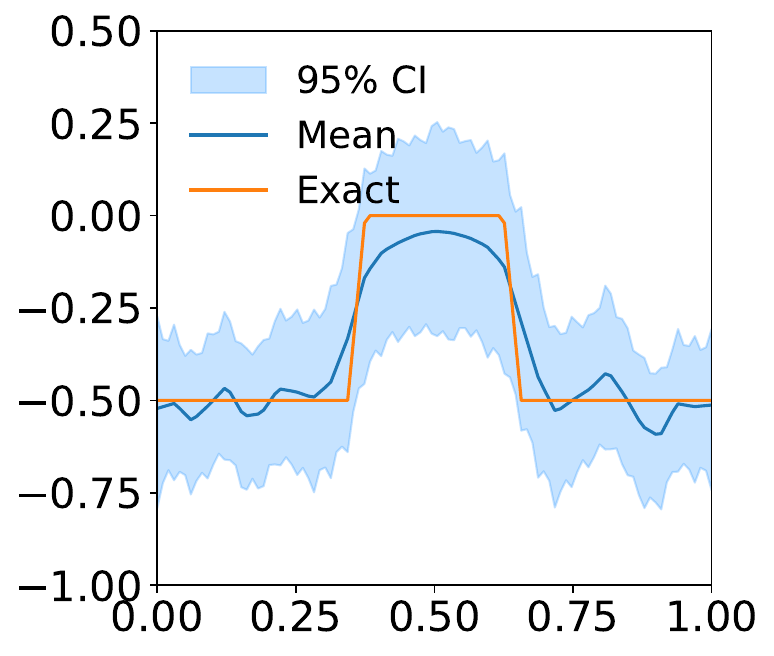 }
    \label{fig:pde-my-ula-post-at-line}}
    \caption{Inference of conductivity $\vecd{\kappa}$ with TV prior through MYULA. $\vecd{\kappa}$ is inferred and plotted on a logarithmic scale.}
    \label{fig:pde-my-ula-full}%
\end{figure}


The exact conductivity is chosen to be a piecewise-constant field on a unit square domain with uniform background and a square inclusion of higher values in the center of the domain, i.e.,
\begin{equation}
    \kappa_{\text{true}}(x_1, x_2) = \begin{cases}
        1,\quad &\text{if } 0.35\leq x_1 < 0.65 \text{ and } 0.35\leq x_2 < 0.65, \text{ and}\\
        \exp(-0.5),\quad &\text{otherwise,}
    \end{cases}
\end{equation}
as shown in Figure~\ref{fig:pde-my-ula-k-true}. The conductivity is considered and later inferred on a logarithmic scale to ensure its positivity, as in \cite{chaudhry2025effects, ernst2024uncertainty}, and the exponential transformation is integrated as part of the forward operator $\lopd{A}_{\vecd{\kappa}}$.

The two-dimensional unit square domain is partitioned into a uniform mesh of 32$\times$32$\times$2 triangular elements. A homogenous Dirichlet boundary condition is applied over the whole domain, i.e., $g=0$, and the source term $f$ is set to be constant as 1. The relative noise level, which we define as $||\vecd{e}||_2/ ||\vecd{y}||_2$, is set to be approximately $1\%$. The exact and noisy data generated with ground truth is shown in Figure~\ref{fig:pde-my-ula-u-exact} and Figure~\ref{fig:pde-my-ula-y-obs}, respectively.


As to the implicit prior, we use the TV-Chambolle denoiser from the \codefont{scikit-image} library \cite{scikit-image} as a proximal operator in the MYULA framework \eqref{eq:my_ula}, similar to Subsection~\ref{sec:tv_prior_cuqipy}. We choose the \code{weight} parameter for the \code{denoise\_tv\_chambolle} to be $0.5\sigma_{e}^2\omega$ where $\omega = 5$. This \code{weight} parameter controls the strength of the TV regularization, with a larger value leading to a stronger regularization.

The posterior sampling is performed with MYULA, with a step size of $10^{-4}$. A total of 500,000 samples are generated, and online thinning is applied to reduce the computational burden and the correlation between samples. As shown in Figure~\ref{fig:pde-my-ula-post-mean}, the posterior mean reasonably captures the qualitative and quantitative features of the background and the square-shaped inclusion. The pointwise standard deviation, Figure~\ref{fig:pde-my-ula-post-std}, shows that the reconstruction generally has slightly higher uncertainty at the inclusion. The $95\%$ credible interval of the posterior through the center of the domain, Figure~\ref{fig:pde-my-ula-post-at-line}, in the horizontal direction shows a similar trend.

\begin{figure}[tb]
    \centering
    \subfloat[$\omega=1.0$]{\includegraphics[width=0.32\linewidth]{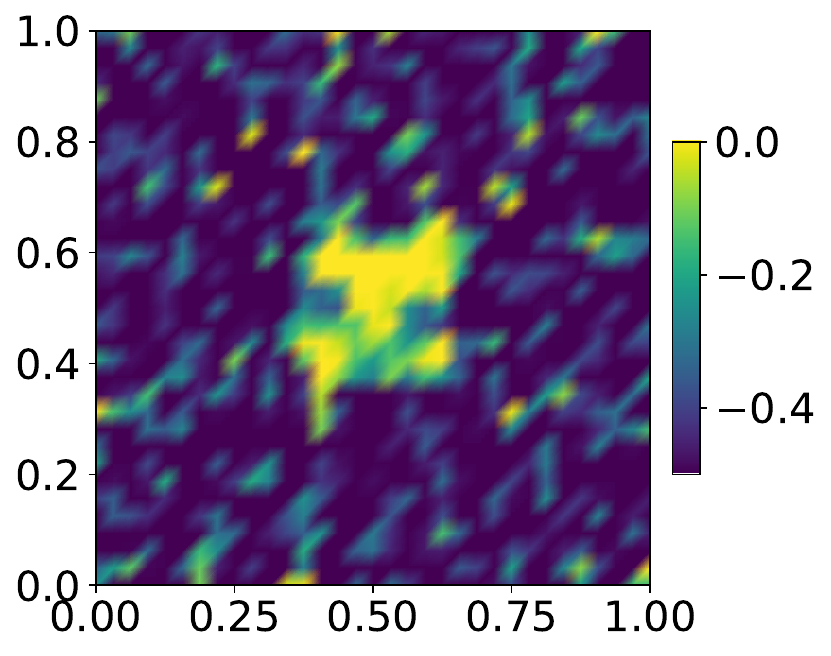} \label{fig:pde-my-ula-reg-stength-1}}
    \hfill
    \subfloat[$\omega=5.0$]{\includegraphics[width=0.32\linewidth]{figures/pdes_myula/mean_5.pdf} 
    \label{fig:pde-my-ula-reg-stength-5}}%
    \hfill
    \subfloat[$\omega=7.0$]{\includegraphics[width=0.32\linewidth]{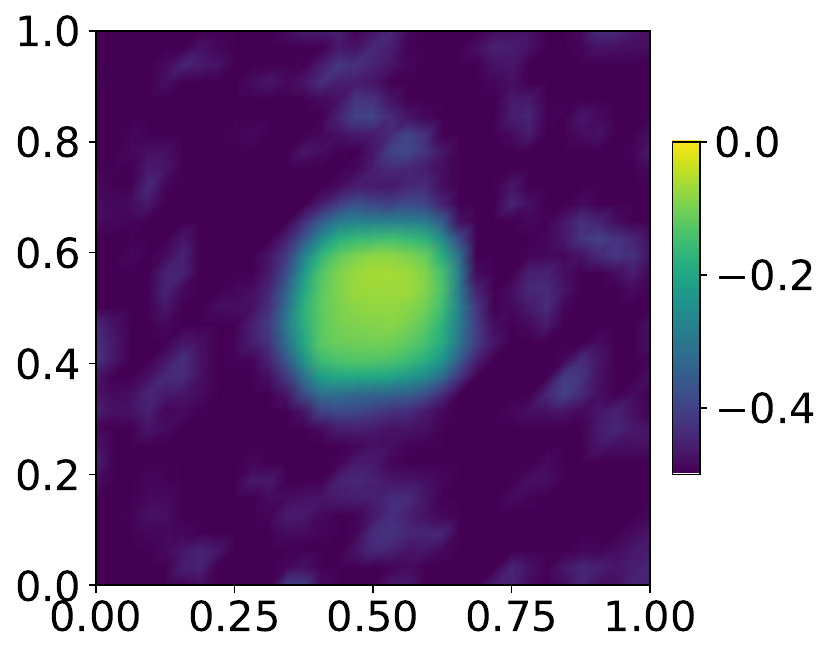} 
    \label{fig:pde-my-ula-reg-stength-7}}
    \\
    \subfloat[$\omega=10.0$]{\includegraphics[width=0.32\linewidth]{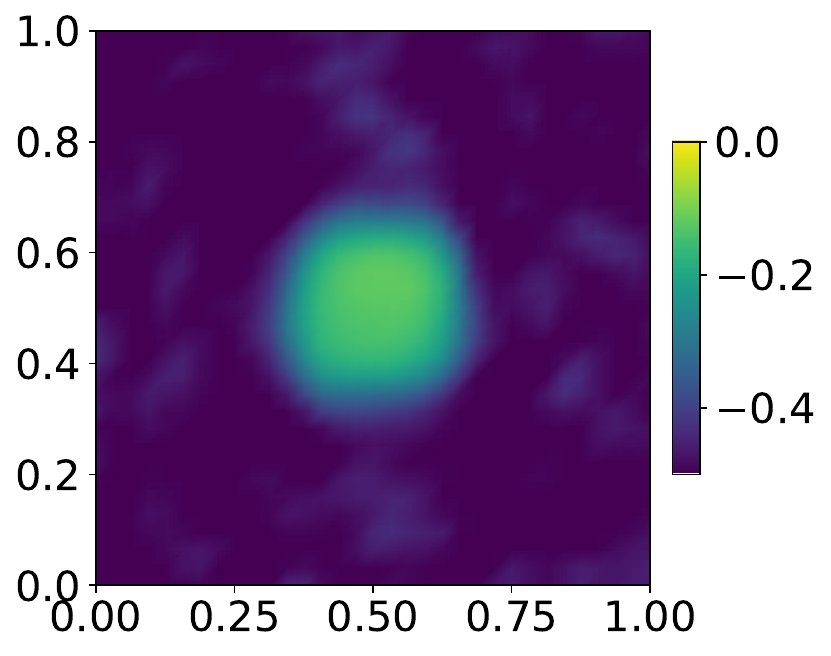}
    \label{fig:pde-my-ula-reg-stength-10}}
    \hfill
    \subfloat[$\omega=20.0$]{\includegraphics[width=0.32\linewidth]{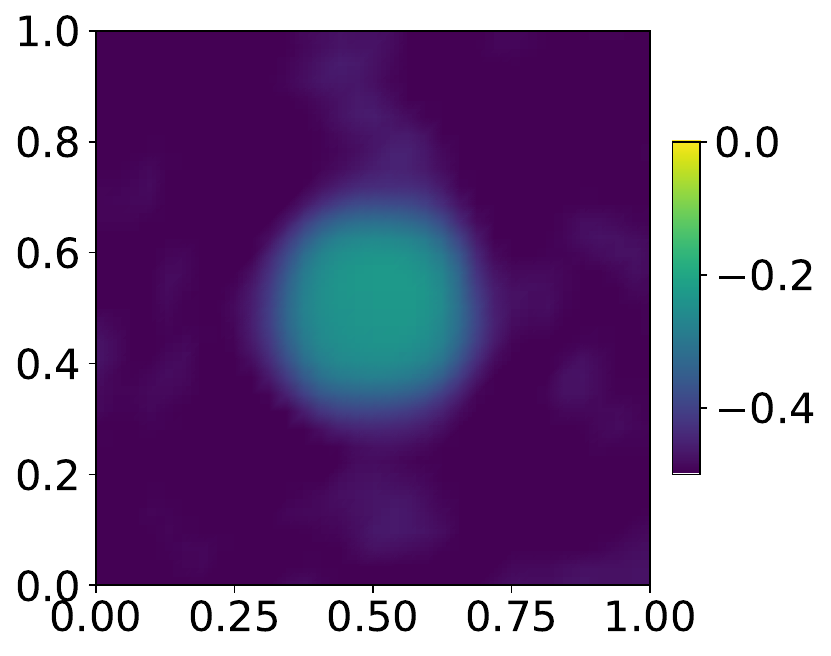}
    \label{fig:pde-my-ula-reg-stength-20}}%
    \hfill
    \subfloat[$\omega=30.0$]{\includegraphics[width=0.32\linewidth]{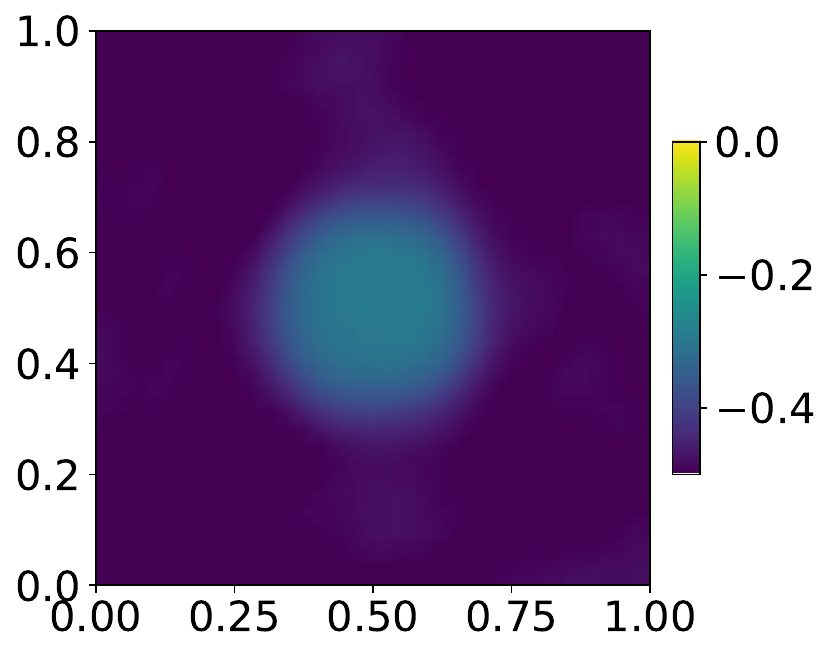}
    \label{fig:pde-my-ula-reg-stength-30}}
    \caption{Posterior means of $\vecd{\kappa}$ with different values of TV weight factor. $\vecd{\kappa}$ is inferred and plotted on a logarithmic scale.}%
    \label{fig:pde-my-ula-diff-reg-strength}%
\end{figure}

We, additionally, experiment with various values of the factor $\omega$ to study the effect of the regularization strength on the inference, and the posterior means under each case are plotted in Figure~\ref{fig:pde-my-ula-diff-reg-strength}. A lower value of $\omega$ leads to a reconstruction with significant oscillation, Figure~\ref{fig:pde-my-ula-reg-stength-1}. As we increase $\omega$, the corresponding reconstruction becomes closer to a piecewise-constant function and still captures the square shape of the inclusion, Figures~\ref{fig:pde-my-ula-reg-stength-5}--\ref{fig:pde-my-ula-reg-stength-10}. For higher values of $\omega$, the inclusions become rounder and their magnitude diminishes, Figures~\ref{fig:pde-my-ula-reg-stength-20} and \ref{fig:pde-my-ula-reg-stength-30}. 

\subsection{Image inpainting with model-based and data-based implicit priors}

This experiment demonstrates the use of implicit priors with PnP-ULA in solving the image inpainting task. Image inpainting refers to reconstructing missing or corrupted regions in images and it is often formulated as an inverse problem where the goal is to recover the original image from partial, often noisy observations. The problem is inherently ill-posed owing to its underdetermined nature, with fewer measurements than unknowns, necessitating regularization through prior information. To regularize the ill-posedness, this experiment explores the use of two types of priors: a wavelet-based thresholding denoiser \cite{chang2000adaptive} of Haar type from \codefont{scikit-image} \cite{scikit-image} with MYULA and a denoising convolutional neural network (DnCNN) denoiser \cite{pesquet2021learning} from \codefont{DeepInverse} \cite{tachella2025deepinverse} with PnP-ULA.



\begin{figure}[tb]
    \centering
    \subfloat[\centering Ground truth]{{\includegraphics[width=0.2\linewidth]{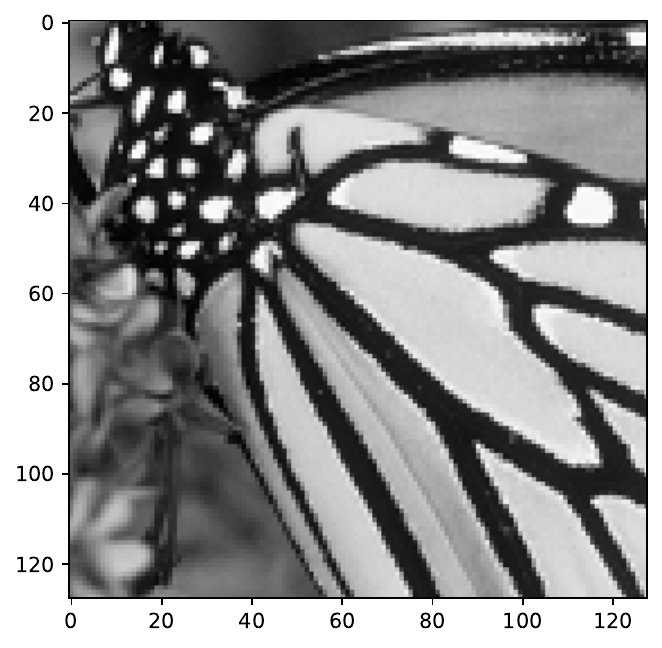} }\label{fig:inpainting-groundtruth}}%
    \subfloat[\centering Mask]{{\includegraphics[width=0.2\linewidth]{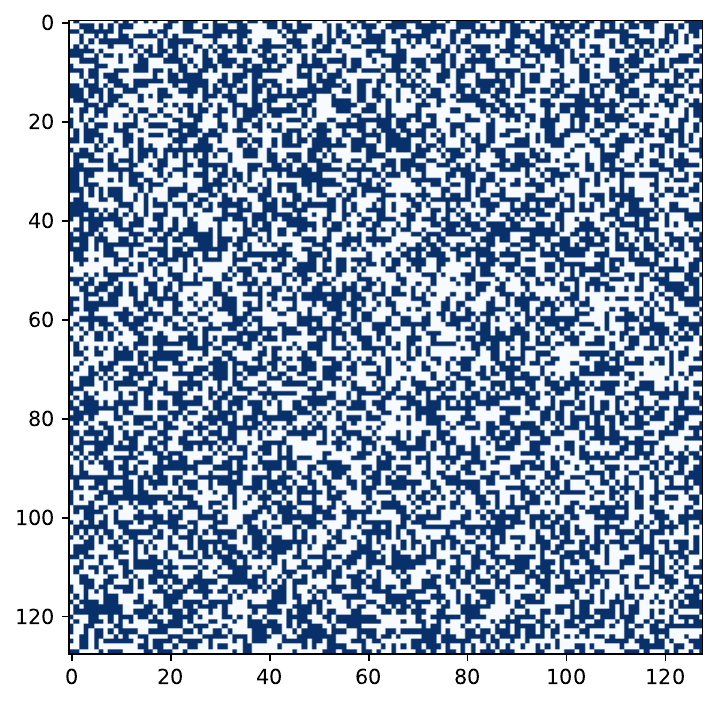} }\label{fig:inpainting-mask}}
    \subfloat[\centering Data]{{\includegraphics[width=0.2\linewidth]{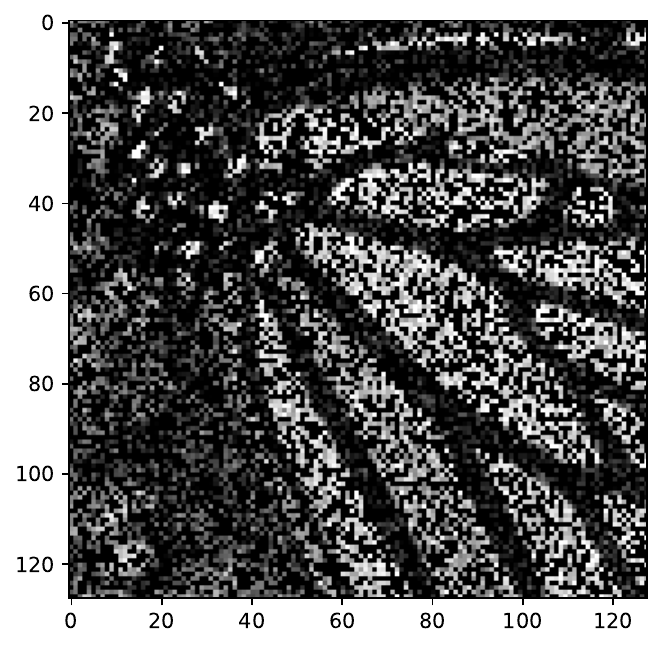} }\label{fig:inpainting-data}}
    \\
    \subfloat[\centering Mean (wavelet)]{{\includegraphics[width=0.2\linewidth]{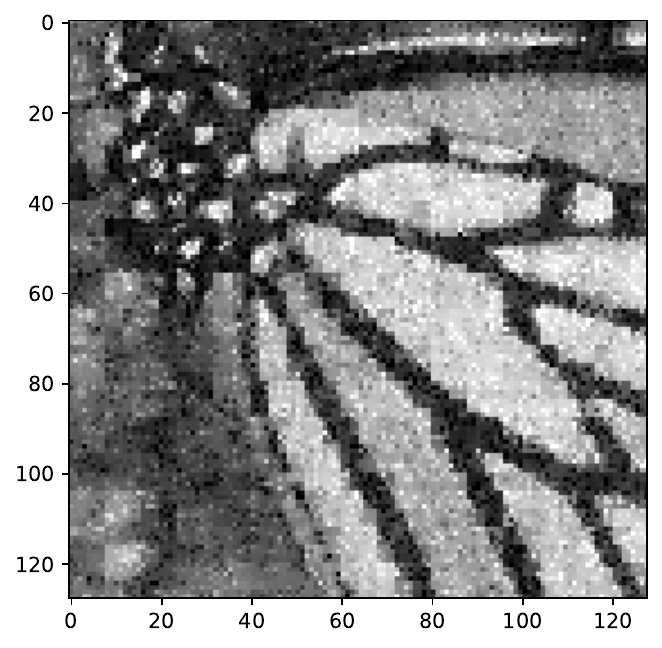} }\label{fig:inpainting-wavelet-mean}}
    \subfloat[\centering 95\% CI (wavelet)]{{\includegraphics[width=0.24\linewidth]{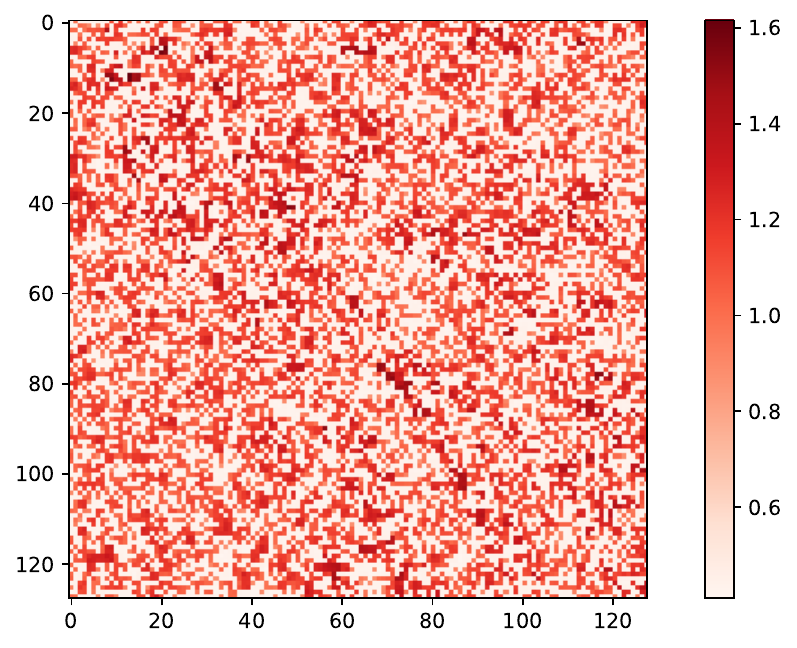} }\label{fig:inpainting-wavelet-std}}
    \subfloat[\centering Mean (DnCNN)]{{\includegraphics[width=0.2\linewidth]{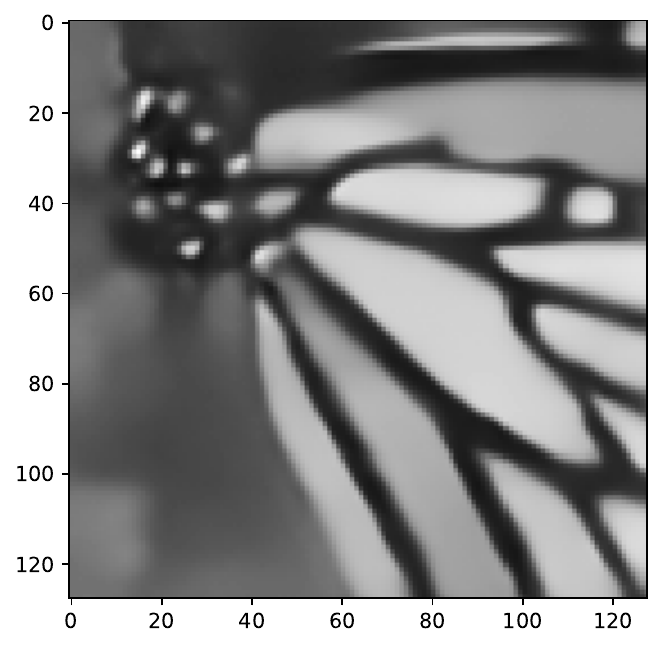} }\label{fig:inpainting-dncnn-mean}}
    \subfloat[\centering 95\% CI (DnCNN)]{{\includegraphics[width=0.245\linewidth]{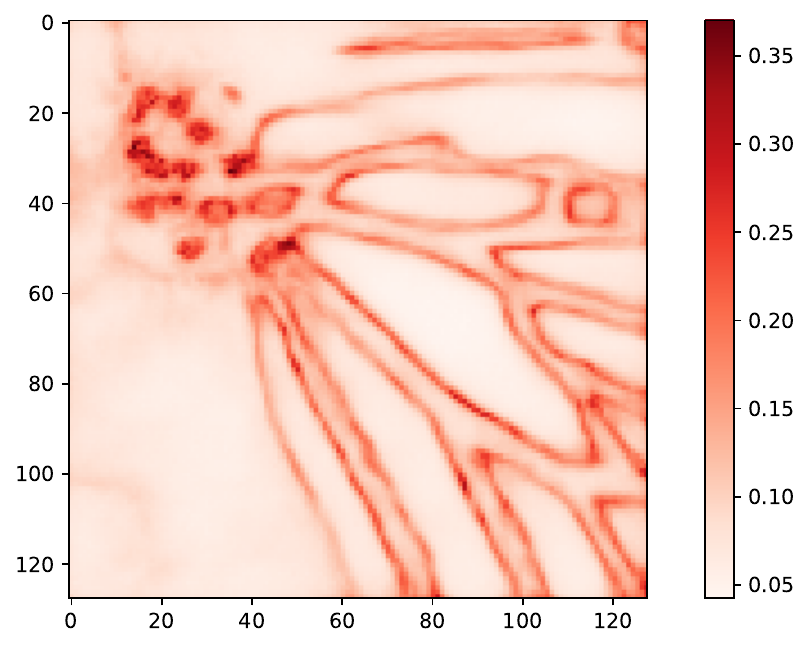} }\label{fig:inpainting-dncnn-std}}
    \caption{Results for the image inpainting problem. The first row shows the ground truth image, the pattern of the binary mask operator (blue indicates removed pixels), and the measured data. The second row shows the means and the 95\% CI width of the posterior samples with the wavelet and the DnCNN priors.}%
    \label{fig:inpainting}%
\end{figure}

In this experiment, a grayscale image is first loaded and resized to a size of 128$\times$128 pixels as the ground truth (Figure~\ref{fig:inpainting-groundtruth}). A linear forward operator representing a binary mask is applied to simulate missing pixel values by randomly removing 50\% of the image pixels, shown in Figure~\ref{fig:inpainting-mask}. The unmasked pixels are further corrupted with additive Gaussian noise of standard deviation 0.1, as shown in Figure~\ref{fig:inpainting-data}. 



With both the wavelet and the DnCNN denoisers, 100,000 samples are generated. Their first 40,000 are discarded as burn-in to ensure convergence. Posterior statistics, including the mean and the standard deviation, are computed from the remaining samples with a thinning factor of 10 to reduce autocorrelation.

As shown in Figure~\ref{fig:inpainting-wavelet-mean} and \ref{fig:inpainting-dncnn-mean}, both of the posterior means obtained with the wavelet and the DnCNN priors exhibit substantial improvement in visual quality compared to the observed data, with the wavelet prior’s mean exhibiting more oscillations than the smoother DnCNN prior’s mean. The 95\% CI of the wavelet prior samples, Figure~\ref{fig:inpainting-wavelet-std}, showing significant similarity with the masking pattern (Figure~\ref{fig:inpainting-mask}), is generally higher at pixels where the mask operator is applied. In contrast, DnCNN prior samples, Figure~\ref{fig:inpainting-dncnn-std}, show higher 95\% CI width near sharp edges in the ground truth, indicating better edge preservation.



This experiment demonstrates the use of implicit priors in the PnP-ULA framework, and shows also the flexibility of \codefont{CUQIpy} in integrating various state-of-the-art denoisers from external libraries, such as \codefont{scikit-image} and \codefont{DeepInverse}, with minimal implementation overhead. We note that better results could be achieved with more appropriate hyperparameters. However, it is not the scope of this paper. We refer to \cite{laumont2022bayesian} and \cite{laumont2024sampling} for more in-depth analysis of the PnP-ULA framework for image inpainting.



\section{Conclusion}\label{sec:conclusion}

In this work, we presented a unified, computational framework for examining the abstract concept of implicit priors for Bayesian inverse problems that contains various priors commonly used and studied in the literature. We discussed the difficulty in distinguishing between explicit and implicit priors, due to some of these priors being able to be transformed into each other with varying difficulty. Then, we discussed how to incorporate implicit priors within a computational Bayesian framework and the dilemma caused by some implicit prior information being introduced in the sampling stage. We presented an implementation for various implicit priors within the \cuqipy\,software package, supported by two simple two-dimensional Bayesian inverse problems. To highlight various use cases for implicit priors and their flexibility, we applied the discussed methods to various commonly occurring inverse problems.

The unification of implicit priors naturally gives rise to new implicit priors which could be implemented and studied in the future, e.g., PnP-RLRTO and PnP-PGLA. Furthermore, some of the mentioned priors are more prominently used in one field than another, with PnP in particular being popular within the imaging community. By making the methods more easily accessible, we hope that they can be applied to many more problems in the future.

\section*{Acknowledgements}
This work was supported by The Villum Foundation (Grant No. 25893).

\bibliographystyle{plain}
\bibliography{references} 


\end{document}